\documentclass[a4paper,11pt]{article}
\pdfoutput=1 

\usepackage{jheppub}

\usepackage[T1]{fontenc} 

\usepackage{amssymb}
\usepackage{hyperref}
\usepackage{latexsym}
\usepackage{bm}
\usepackage{slashed}
\usepackage{datetime}
\usepackage{mciteplus}
\usepackage{multirow}
\usepackage{color}
\usepackage[utf8]{inputenc}
\usepackage[normalem]{ulem}

\usepackage{pdflscape} 
\usepackage{makecell}  

\usepackage[section]{placeins}

\newcommand{\T}{\perp}

\newcommand{\bT}{b_T}
\newcommand{\bb}{b}
\newcommand{\bmin}{\bb_{\rm min}}
\newcommand{\bmax}{\bb_{\rm max}}
\newcommand{\bstar}{\bb_*}
\newcommand{\F}{\hat{f}_1}

\allowdisplaybreaks[2]

\title{Transverse-momentum-dependent parton distributions up to N$^3$LL from Drell-Yan data}

\author[a,b]{Alessandro Bacchetta,}
\author[a,b]{Valerio Bertone,}
\author[a,b]{Chiara Bissolotti,}
\author[a,b]{Giuseppe Bozzi,}
\author[c]{Filippo Delcarro,}
\author[a,b]{Fulvio Piacenza,}
\author[b]{and Marco Radici}

\affiliation[a]{Dipartimento di Fisica, Universit\`a di Pavia, via Bassi
  6, I-27100 Pavia}

\affiliation[b]{INFN Sezione di Pavia, via Bassi 6,
  I-27100 Pavia, Italy}

\affiliation[c]{Jefferson Lab, 12000
  Jefferson Avenue, Newport News, Virginia 23606, USA}

\emailAdd{alessandro.bacchetta@unipv.it}
\emailAdd{valerio.bertone@cern.ch}
\emailAdd{chiara.bissolotti01@universitadipavia.it}
\emailAdd{giuseppe.bozzi@unipv.it}
\emailAdd{delcarro@jlab.org}
\emailAdd{fulvio.piacenza01@universitadipavia.it}
\emailAdd{marco.radici@pv.infn.it}

\preprint{JLAB-THY-19-3121}


\abstract {We present an extraction of unpolarised
  Transverse-Momentum-Dependent Parton Distribution Functions
  based on Drell-Yan production data from different
  experiments, including those at the LHC, and spanning a wide
  kinematic range. We deal with experimental uncertainties by properly taking
  into account correlations. We include resummation of logarithms of the
  transverse momentum of the vector boson up to N$^3$LL order,
  and we include non-perturbative contributions. These ingredients allow
  us to obtain a remarkable agreement with the data.
}

\begin{document}
\maketitle
\flushbottom

\section{Introduction}
\label{s:intro}

The analysis of hard scattering processes involving nucleons in the
initial state allows us to obtain information on their internal
structure, encoded in parton distribution functions (PDFs).

After decades of studies, we have obtained a detailed knowledge of
unpolarised collinear PDFs: they provide information about matter at
the subnuclear level and are indispensable in almost any prediction
involving high-energy hadrons.  Collinear PDFs describe the
distribution of partons inside the nucleon as a function of the
longitudinal momentum fraction $x$.  Collinear factorisation theorems
lead to a precise definition of collinear PDFs based on perturbative
QCD and, within specific approximations, determine also their
connection to experimental observables.

When considering semi-inclusive observables, factorisation theorems
require the introduction of more general PDFs. We will focus in
particular on the $q_T$ distribution of vector bosons ($\gamma$ and
$Z$) produced in Drell-Yan processes. At low $q_T$, this observable
can be written in terms of Transverse-Momentum-Dependent Parton
Distribution Functions (TMD PDFs or, in short, TMDs), which describe
the distribution of partons as a function not only of the longitudinal
momentum fraction $x$, but also on the partonic transverse momentum
$k_\T$ (see, \textit{e.g.},
Refs.~\cite{Rogers:2015sqa,Diehl:2015uka,Angeles-Martinez:2015sea} and
references therein).  TMDs are partially computable by means of
well-established perturbative methods that take into account soft and
collinear radiation to all orders. However, calculations based on
perturbative QCD become unreliable for values of transverse momentum
close to the Landau pole ($\Lambda_{\mathrm{QCD}}$). In this regime,
non-perturbative components have to be included and have to be
determined through fits to experimental data.

Several works in the past have studied the non-perturbative components
in Drell-Yan $q_T$ distributions~\cite{Davies:1984sp,Ladinsky:1993zn,
  Landry:1999an,Qiu:2000hf,Landry:2002ix,Konychev:2005iy,
  Becher:2011xn,Camarda:2019zyx} or in semi-inclusive
DIS~\cite{Meng:1995yn,Nadolsky:1999kb},
without directly mentioning
TMDs.
More recent
works directly performed extractions of TMDs from Drell-Yan
data~\cite{DAlesio:2014mrz,Scimemi:2017etj,Bertone:2019nxa},
semi-inclusive DIS data~\cite{Signori:2013mda,Anselmino:2013lza} or
both~\cite{Echevarria:2014xaa,Su:2014wpa,Bacchetta:2017gcc,Scimemi:2019cmh}. 
Alternatively, TMDs were determined in the so-called
parton-branching approach by
solving evolution equations with an iterative method similar to parton showers
but including transverse momentum
dependence~\cite{Martinez:2018jxt,Martinez:2019mwt}. 

A precise knowledge of TMDs if useful not only to investigate the
structure of the nucleon in greater detail, but also to improve the
reliability of predictions involving TMDs. At high energies, the
perturbative part of TMDs may be dominant, but when extreme precision
is required, also the non-perturbative components become relevant
(see, \textit{e.g.}, Ref.~\cite{Bacchetta:2018lna}).

In this work, we will determine the unpolarised quark TMDs by fitting
Drell-Yan data from experiments at Tevatron, RHIC, LHC, and low-energy
experiments at Fermilab, for a total of around 350 data points.  The
dataset is similar to the one studied in Ref.~\cite{Bertone:2019nxa},
but there are some important differences: whenever available, we use
cross-section measurements without any normalisation factor; TMD
evolution is implemented in a different way; for
the first time, TMD evolution is implemented up to
next-to-next-to-next-to-leading logarithmic (N$^3$LL) accuracy.
Compared to Ref.~\cite{Bacchetta:2017gcc}, we exclude data from
semi-inclusive Deep Inelastic Scattering, but we greatly extend the
Drell-Yan data dataset, we improve the logarithmic accuracy, we study
normalisations with much greater care, and we abandon the narrow-width
approximation for $Z$ -boson production data.

The paper is organised as follows. In Sec.~\ref{s:theory}, we give
some details of the theoretical framework. In Sec.~\ref{s:data}, we
describe the selection of experimental data. In Sec.~\ref{s:results},
we show our results.  Finally, in Sec.~\ref{s:conc} we draw our
conclusions.

\section{Theoretical framework} \label{s:theory}

In this section we describe the theoretical framework of our analysis.
In Sec.~\ref{s:DYxsec}, we review the TMD factorisation formula for
the Drell-Yan (DY) process. In Sec.~\ref{s:evmatching}, we briefly
describe the evolution of TMDs and how they can be matched onto the
collinear PDFs. Sec.~\ref{s:pertcont} collects the perturbative
ingredients of the factorised formula within the particular choice of
the evolution scales adopted in this analysis. In
Sec.~\ref{s:pertordering}, we discuss how these perturbative
ingredients are to be combined to achieve a given logarithmic accuracy
of the resummation provided by TMD factorisation. In this context, we
also review the different logarithmic-counting prescriptions used in
the literature, highlighting the possible differences. Finally, in
Sec.~\ref{s:npfunc} we motivate the introduction of a non-perturbative
contribution that needs to be determined from data, and we discuss its
particular functional form.

\subsection{Drell-Yan cross section in TMD factorisation}
\label{s:DYxsec}

\begin{figure}[tbp]
\centering
\includegraphics[width=0.8\textwidth]{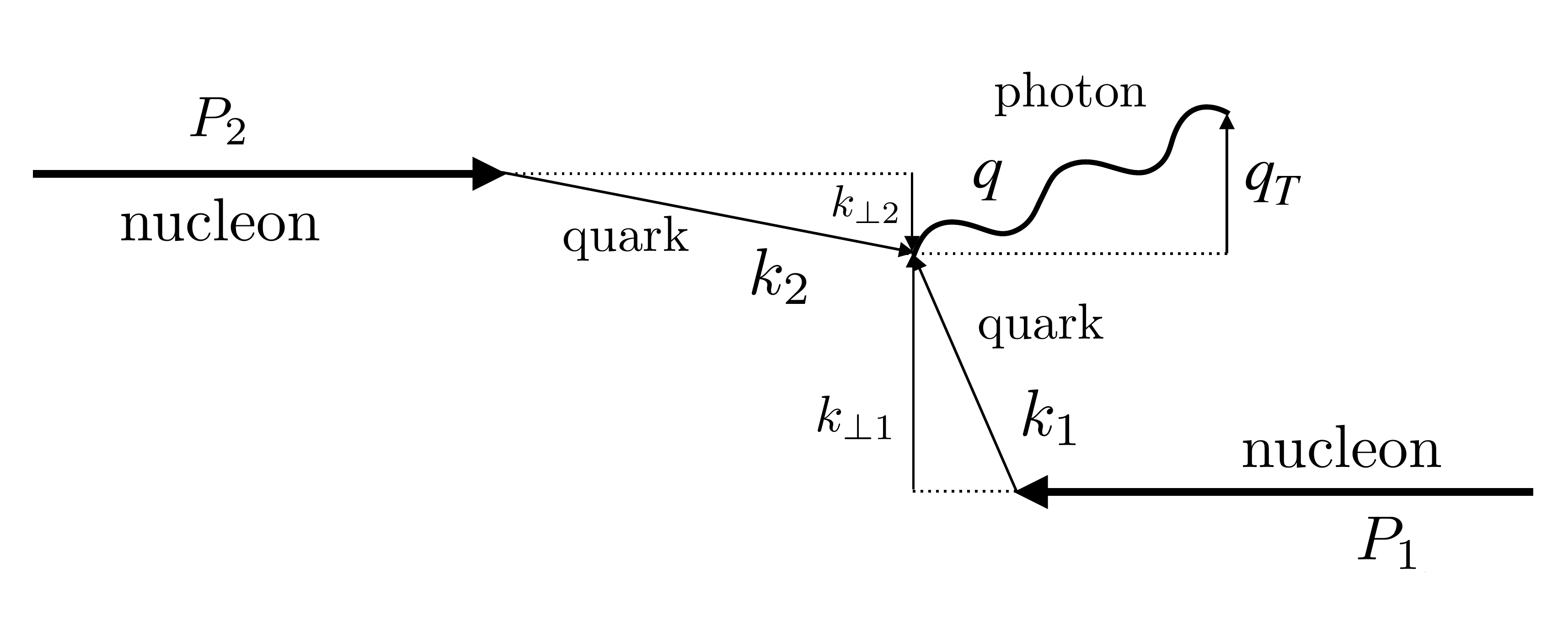}
\caption{Diagram displaying the relevant momenta involved in a
  Drell-Yan event. In a reference frame in which two colliding
  nucleons move along the $z$ direction with 4-momenta $P_1$ and
  $P_2$, a quark with 4-momentum $k_1$ and transverse momentum
  $\bm{k}_{\T 1}$ annihilates with a parton with 4-momentum $k_2$ and
  transverse momentum $\bm{k}_{\T 2}$. A (virtual) photon (or $Z$) is
  produced with 4-momentum $q$ and transverse momentum
  $\bm{q}_T=\bm{k}_{\T 1}+\bm{k}_{\T 2}$ .}
\label{f:trans_momenta_DY}
\end{figure}
In the inclusive Drell-Yan process
\begin{equation}
  h_1(P_1)+h_2(P_2)\longrightarrow \gamma^*/Z (q) + X \longrightarrow \ell^+(l) + \ell^-(l') + X \; ,
\end{equation}
two hadrons $h_1$ and $h_2$ with 4-momenta $P_1$ and $P_2$,
respectively, collide with center-of-mass energy squared
$s = (P_1+P_2)^2$ and produce a neutral vector boson $\gamma^*/Z$ with
4-momentum $q$ and large invariant mass $Q=\sqrt{q^2}$. The vector
boson eventually decays into a lepton and an antilepton with 4-momenta
constrained by momentum conservation, $q = l + l'$.  The absolute
value of the transverse momentum and the rapidity of the neutral boson
(or, equivalently, of the lepton pair) are defined as
\begin{equation}
  \label{eq:qTrapidity}
  q_T = \sqrt{q_x^2+q_y^2} \;, \qquad y = \frac12 \ln \left( \frac{q_0+q_z}{q_0-q_z} \right) \;,
\end{equation}
where the $z$ direction is defined by the hadronic-collision axis (see
Fig.~\ref{f:trans_momenta_DY}).

We are specifically interested in the transverse-momentum distribution
of the vector boson in the small-$q_T$ region ($q_T \ll Q$).  In this
regime, the (unpolarised) differential cross section factorises and
can be expressed in terms of the (unpolarised) TMDs of the two hadrons
as
\begin{equation}
  \label{eq:crosssectionkT}
  \begin{split}
  &\frac{d\sigma}{dQ dy dq_T} =
  \frac{16\pi^2\alpha^2q_T\mathcal{P}}{9 Q^3} H(Q,\mu) \, \sum_q c_q(Q)
  \\
  &\quad \times \int d^2\bm{k}_{\T 1}^{} \, d^2\bm{k}_{\T 2}^{} \,  x_1 f_1^q \big( x_1,\bm{k}_{\T 1}^2;\mu,\zeta_1 \big)
\, x_2 f_{1}^{\bar{q}} \big( x_2,\bm{k}_{\T 2}^2; \mu,\zeta_2 \big) \, \delta^{(2)} \big( {\bm k}_{\T 1} + {\bm k}_{\T 2} - {\bm q}_T \big) \; ,
\end{split}
\end{equation}
where $\alpha$ is the electromagnetic coupling and $\mathcal{P}$ is
the phase-space reduction factor due to possible kinematic cuts on the
final-state leptons (see Appendix~\ref{a:LeptonCuts}).\footnote{In the
  presence of cuts on single lepton variables, an additional
  parity-violating term contributes to the cross
  section~\cite{Boer:1999mm}. However, in Appendix~\ref{a:LeptonCuts}
  we argue that this contribution is negligible in the experimental
  conditions considered in this paper.} The hard factor $H$ represents
the perturbative part of the hard scattering and depends on the hard
scale $Q$ and on the renormalisation scale $\mu$. The summation over
$q$ in Eq.~(\ref{eq:crosssectionkT}) runs over the active quarks and
antiquarks at the scale $Q$, and $c_q$ are the respective electroweak
charges given by
\begin{equation}
\label{eq:fullcoup}
c_q(Q) = e_q^2 - 2 e_q V_q V_\ell \, \chi_1(Q) + (V_\ell^2 + A_\ell^2)\, (V_q^2 + A_q^2)\, \chi_2(Q)\; ,
\end{equation}
with
\begin{align}
\chi_1(Q) &= \frac{1}{4 \sin^2\theta_W \cos^2\theta_W } \frac{Q^2 ( Q^2 -  M_Z^2 )}{ (Q^2 - M_Z^2)^2 + M_Z^2 \Gamma_Z^2} \; ,\\
\chi_2(Q) &= \frac{1}{16 \sin^4\theta_W\cos^4\theta_W} \frac{Q^4}{ (Q^2 - M_Z^2)^2 + M_Z^2 \Gamma_Z^2} \; ,
\end{align}
where $e_q$, $V_q$, and $A_q$ are respectively the electric, vector,
and axial charges of the flavour $q$; $V_\ell$ and $A_\ell$ are the
vector and axial charges of the lepton $\ell$; $\sin\theta_W$ is the
weak mixing angle; $M_Z$ and $\Gamma_Z$ are mass and width of the $Z$
boson.

The second line of Eq.~(\ref{eq:crosssectionkT}) displays the
convolution of the TMDs $f_1^q$ and $f_{1}^{\bar{q}}$ of the hadrons
$h_1$ and $h_2$, respectively. It describes the annihilation of a
quark $q$, with longitudinal momentum fraction
$x_1 = Q e^y / \sqrt{s}$ and transverse momentum $\bm{k}_{\T 1}$, with
the corresponding antiquark $\bar{q}$, with longitudinal momentum
fraction $x_2 = Q e^{-y} / \sqrt{s}$ and transverse momentum
$\bm{k}_{\T 2}$.  In the annihilation, the momentum conservation is
guaranteed by the presence of
$\delta^{(2)} \big( {\bm k}_{\T 1} + {\bm k}_{\T 2} - {\bm q}_T \big)$
(see Fig.~\ref{f:trans_momenta_DY}).

As a consequence of renormalisation and of the removal of the rapidity
divergences~\cite{Collins:2011zzd}, TMDs acquire a dependence on the
renormalisation scale $\mu$ and on the so-called rapidity scale
$\zeta$. We will discuss our choice for these scales in
Sec.~\ref{s:pertcont}. Here, we just remark that the rapidity scales
$\zeta_1$ and $\zeta_2$ in Eq.~(\ref{eq:crosssectionkT}) must obey the
kinematic constraint $\zeta_1 \zeta_2 = Q^4$.

It is convenient to rewrite the convolution in the conjugate position
space by using the Fourier transform of each TMD, defined
as\footnote{For simplicity, in the rest of the paper we will refer to
  the $\bT$-dependent function $\F$ as to TMD but understanding that
  this is in fact the Fourier transform of the actual TMD ${f}_1$.
  Note that in Ref.~\cite{Bacchetta:2017gcc} the variable $\xi_T$ was
  used in place of $\bT$. The reason was to avoid confusion with the
  impact parameter used in the GPD literature for which the symbol
  $\bT$ is typically used. In this paper, we decided to use $\bT$ as
  it is more common in the TMD, $q_T$-resummation, and SCET literature
  but keeping in mind that this is \textit{not} the impact parameter
  but the Fourier conjugate variable of $q_T$. Finally, we notice that
  in Ref.~\cite{Bacchetta:2017gcc} the Fourier transform was defined
  with an extra $1/(2 \pi)$ factor.}
\begin{equation}
\label{eq:FTdef}
  \begin{split}
    \F^q \big( x, \bT; \mu, \zeta \big) &= \int d^2 \bm{k}_\T \, e^{i
      \bm{k}_\T \cdot \bm{\bb}_T} \, f_1^q \big( x, \bm{k}_\T^2; \mu,
    \zeta \big) \, ,
 \end{split}
\end{equation}
where $\bT$ is the absolute value of the vector $\mathbf{\bb}_T$
($\bT=|\mathbf{\bb}_T|$). By using Eq.~(\ref{eq:FTdef}), we can
rewrite the convolution of TMDs as
\begin{equation}
\label{eq:FTTMD}
\begin{split}
  & \int d^2\bm{k}_{\T 1}^{} \, d^2\bm{k}_{\T 2}^{} \, x_1 f_1^q \big(
  x_1,\bm{k}_{\T 1}^2;\mu,\zeta_1 \big) \, x_2 f_{1}^{\bar{q}} \big(
  x_2,\bm{k}_{\T 2}^2; \mu,\zeta_2 \big) \, \delta^{(2)} \big( {\bm
    k}_{\T 1} + {\bm k}_{\T 2} - {\bm q}_T \big)
  \\
  &= \int \frac{d^2\bm{\bb}_T}{(2\pi)^2} \, e^{i \bm{\bb}_T\cdot
    \bm{q}_T} \, x_1 \F^q \big( x_1, \bT; \mu, \zeta_1 \big) \, x_2
  \F^{\bar{q}} \big( x_2, \bT; \mu, \zeta_2 \big)
  \\
  &= \frac{1}{2 \pi} \int_0^\infty d\bT \, \bT \, J_0 \big( \bT q_T
  \big) \, x_1 \F^q \big( x_1, \bT; \mu, \zeta_1 \big) \, x_2
  \F^{\bar{q}} \big( x_2, \bT; \mu, \zeta_2 \big) \; ,
\end{split}
\end{equation}
where $J_0$ is the 0-th order Bessel function of the first kind that
has the following integral representation
\begin{equation}
J_0 (x) = \frac{1}{2 \pi} \int_0^{2\pi} d\theta \, e^{i x \cos\theta} \;.
\end{equation}
By inserting Eq.~(\ref{eq:FTTMD}) into the cross section in
Eq.~(\ref{eq:crosssectionkT}), we finally get
\begin{equation}
  \label{eq:crosssection}
  \begin{split}
  &\frac{d\sigma}{dQ dy dq_T} =
  \frac{8 \pi \alpha^2 q_T \mathcal{P}}{9 Q^3} H(Q,\mu)
  \\
  &\quad \times \sum_q c_q(Q) \int_0^\infty d\bT \, \bT \, J_0 \big( \bT q_T \big) \, x_1 \F^q \big( x_1, \bT; \mu, \zeta_1 \big) \,
    x_2 \F^{\bar{q}} \big( x_2, \bT; \mu, \zeta_2 \big) \; ,
\end{split}
\end{equation}
which is the formula actually implemented in our analysis of Drell-Yan data.

\subsection{TMD evolution and matching}
\label{s:evmatching}

In Eq.~(\ref{eq:crosssection}), the dependence of the TMDs
$\F^{q (\bar{q})}$ on the scales $\mu$ and $\zeta$ arises from the
removal of the ultraviolet and rapidity divergences in their operator
definition. Each dependence is controlled by an evolution equation:
\begin{align}
\label{eq:eveqs}
\frac{\partial \ln \F}{\partial \ln \mu} &= \gamma(\mu,\zeta) \;,
&
\displaystyle \frac{\partial \ln \F}{\partial \ln \sqrt{\zeta}} &= K(\mu) \; ,
\end{align}
where $\gamma$ is the anomalous dimension of the Renormalisation Group
(RG) evolution in $\mu$, and $K$ is the anomalous dimension of the
Collins-Soper evolution in $\sqrt{\zeta}$~\cite{Collins:1981uk}.
Notice that, for brevity, we have dropped the flavour index $q$ and
$\bar{q}$. Moreover, since in this section we will only be concerned
with the dependence of $\F$ on the scales $\mu$ and $\zeta$, we will
also temporarily drop the dependence on $x$ and $\bT$. In addition to
the evolution equations in Eq.~(\ref{eq:eveqs}), the rapidity
anomalous dimension $K$ obeys its own RG equation:
\begin{equation}\label{eq:CuspRGE}
\frac{\partial K}{\partial \ln \mu} = - \gamma_K \big( \alpha_s (\mu) \big) \; ,
\end{equation}
where $\gamma_K$ is known as cusp anomalous dimension. Since the
crossed double derivatives of $\F$ must be equal, using
Eqs.~(\ref{eq:eveqs}) and (\ref{eq:CuspRGE}) we also get
\begin{equation}
\frac{\partial \gamma }{\partial \ln \sqrt{\zeta}} = - \gamma_K \big( \alpha_s (\mu) \big) \; .
\end{equation}
Using the point $\zeta = \mu^2$ as a boundary condition, the solution
of this differential equation is
\begin{equation}
\label{eq:GFEv}
\gamma(\mu,\zeta) = \gamma_F \big( \alpha_s (\mu) \big) -
\gamma_K \big( \alpha_s (\mu) \big) \ln \frac{\sqrt{\zeta}}{\mu} \; ,
\end{equation}
where $\gamma_F(\alpha_s(\mu))\equiv\gamma(\mu,\mu^2)$. If the TMD
$\F$ is known at some starting scales $\mu_0$ and $\zeta_0$, the
solution of the evolution equations in Eq.~(\ref{eq:eveqs}) reads
\begin{equation}
\label{eq:solution2}
  \F(\mu,\zeta) = R\bigl[(\mu,\zeta)\leftarrow
    (\mu_0,\zeta_0)\bigr] \F(\mu_0,\zeta_0)\,,
\end{equation}
where the so-called Sudakov form factor $R$ accounts for the
perturbative evolution of $\F$ and it is defined as
\begin{equation}\label{eq:evkernelexp}
  R \bigl[ (\mu,\zeta) \leftarrow
    (\mu_0,\zeta_0) \bigr] = \exp \left\{ K(\mu_0) \ln \frac{\sqrt{\zeta}}{\sqrt{\zeta_0}} + \int_{\mu_0}^{\mu} \frac{d\mu'}{\mu'}\left[ \gamma_F (\alpha_s(\mu')) - \gamma_K (\alpha_s(\mu')) \ln \frac{\sqrt{\zeta}}{\mu'} \right] \right\} \; .
\end{equation}
We note that Eq.~(\ref{eq:evkernelexp})
can be implemented in various
ways~\cite{Chiu:2011qc,Chiu:2012ir,Scimemi:2018xaf,Billis:2019evv}.
In this work, we follow the standard approach described
in~\cite{Collins:2011zzd}. Moreover, we calculate all
ingredients involved in Eq.~(\ref{eq:evkernelexp}) by adopting a fully
numerical approach. 

An important property of the TMD $\F$ is that at small values of $\bT$
it can be matched onto the \textit{collinear} PDF $f_1$. Reinstating
for clarity the $x$ and $\bT$ dependence and introducing the matching
coefficient function $C$, we can write\footnote{A sum over flavours is
  understood. The matching function $C$ has to be regarded as a matrix
  in flavour space multiplying a column vector of collinear PDFs.}
\begin{equation}
\label{eq:matching}
  \F(x,\bT;\mu_0,\zeta_0) = \int_x^1 \frac{dy}{y} C(y,\bT;\mu_0,\zeta_0) f_1
  \bigg( \frac{x}{y}; \mu_0 \bigg) \equiv \bigl[ C \otimes f_1 \bigr](x,\bT;\mu_0,\zeta_0) \; .
\end{equation}
Then, the actual evolved TMD becomes
\begin{equation}
\label{eq:solution3}
  \F(x,\bT;\mu,\zeta) = R \bigl[\bT; (\mu,\zeta) \leftarrow (\mu_0,\zeta_0)
    \bigr] \bigl[ C \otimes f_1 \bigr](x,\bT;\mu_0,\zeta_0)\; .
\end{equation}

\subsection{Perturbative content}
\label{s:pertcont}

In order to use Eq.~(\ref{eq:solution3}) in phenomenological
applications, we need to define the values of both the initial and
final pairs of scales, $(\mu_0,\zeta_0)$ and $(\mu,\zeta)$. It turns
out that in the $\overline{\mbox{MS}}$ renormalisation scheme there
exists a particular scale,
\begin{equation}
\label{eq:mub}
  \mu_b(\bT) = \frac{2e^{-\gamma_E}}{\bT} \; ,
\end{equation}
with $\gamma_E$ the Euler constant, such that the rapidity anomalous
dimension $K$ and the matching coefficient $C$ computed at
$\mu_0=\sqrt{\zeta_0}=\mu_b$ admit a pure perturbative expansion free
of explicit logarithms of the scales. Therefore, $\mu_b$ provides a
natural choice for $\mu_0$ and $\sqrt{\zeta_0}$.

The final renormalisation scale $\mu$ must match the one used in the
hard factor $H$ in Eq.~(\ref{eq:crosssection}). Therefore, $\mu$ has
to be of order $Q$ for avoiding large logarithms in $H$: we choose
$\mu=Q$. Any variation of $\mu$ with respect to this choice can be
accounted for by expanding the solution of the RG equation for the
strong coupling $\alpha_s$. The rapidity scales $\zeta_1$ and $\zeta_2$ in
Eq.~(\ref{eq:crosssection}) are bound to comply with
$\zeta_1\zeta_2 = Q^4$. Therefore, the natural choice is
$\zeta_1=\zeta_2=Q^2$. However, we stress that any choice that fulfils
this constraint leads to the same cross section. In fact, from
Eq.~(\ref{eq:evkernelexp}) it should be evident that the evolution
factors $R$ entering the two TMDs in Eq.~(\ref{eq:crosssection})
combine in such a way that the result only depends on the product
$\zeta_1\zeta_2$.

After choosing the scales, we discuss the perturbative ingredients
that result from this particular choice. We first consider the hard
function $H$. Up to two-loop accuracy, its perturbative expansion is
\begin{equation}
\label{eq:Hexp}
H(Q,Q) = 1 + \sum_{n=1}^{2} \left( \frac{\alpha_s(Q)}{4\pi} \right)^n H^{(n)} \; .
\end{equation}
The coefficients $H^{(n)}$ can be read off from, \textit{e.g}.,
Ref.~\cite{Bizon:2018foh}. When going beyond
$\mathcal{O}(\alpha_s^2)$, the hard function acquires a non-trivial
flavour structure (see, \textit{e.g.},
Ref.~\cite{Collins:2017oxh}). As a consequence, $H$ should in
principle be moved inside the flavour sum in
Eq.~(\ref{eq:crosssection}). However, in the present analysis we do
not consider corrections beyond $\mathcal{O}(\alpha_s^2)$ and
Eq.~(\ref{eq:crosssection}) is appropriate.

Next, we consider the matching function $C$ introduced in
Eq.~(\ref{eq:matching}). By making the flavour and $x$ dependences
explicit, the $C$ have the following perturbative expansion
\begin{equation}
\label{eq:Cexp}
  C_{ij}(x,\bT;\mu_b,\mu_b^2) = \delta_{ij} \delta(1-x) + \sum_{n=1}^{\infty} \left( \frac{\alpha_s(\mu_b)}{4\pi} \right)^n C_{ij}^{(n)}(x) \; .
\end{equation}
The coefficient functions $C_{ij}^{(n)}$ up to $n=2$ have been
computed in Refs.~\cite{Catani:2012qa,Echevarria:2016scs}. They have
been reported also in Ref.~\cite{Collins:2017oxh}, where the authors
have verified the consistency of the results. The calculation of the
$\mathcal{O}(\alpha_s^3)$ corrections to the quark matching functions
appeared very recently in Ref.~\cite{Luo:2019szz}.

As for the anomalous dimensions $K$, $\gamma_F$, and $\gamma_K$ in the
Sudakov form factor in Eq.~(\ref{eq:evkernelexp}), their perturbative
expansions read, respectively,
\begin{equation}
\label{eq:andimexp}
\begin{array}{rcl}
K(\mu_b) &=&\displaystyle
             \sum_{n=0}^{\infty} \left( \frac{\alpha_s(\mu_b)}{4\pi} \right)^{n+1}
             K^{(n)} \;, \\
\\
\gamma_F(\alpha_s(\mu)) &=&\displaystyle
             \sum_{n=0}^{\infty} \left( \frac{\alpha_s(\mu)}{4\pi} \right)^{n+1}
             \gamma_F^{(n)} \;, \\
\\
\gamma_K(\alpha_s(\mu)) &=&\displaystyle
             \sum_{n=0}^{\infty} \left( \frac{\alpha_s(\mu)}{4\pi}\right)^{n+1}
             \gamma_K^{(n)} \;.
\end{array}
\end{equation}
The coefficients $K^{(n)}$ are listed up to $n=3$ in
Ref.~\cite{Echevarria:2016scs} and up to $n=2$ in
Ref.~\cite{Collins:2017oxh}. They differ by a factor $-2$ due to a
different definition of $K$. Also the coefficients $\gamma_F^{(n)}$
are given in Refs.~\cite{Collins:2017oxh,Echevarria:2016scs} up to
$n=2$, and they differ by a minus sign due to a different definition
of the anomalous dimension. Finally, the coefficients
$\gamma_K^{(n)}$ were originally computed in Ref.~\cite{Li:2016ctv}
and are also given in
Refs.~\cite{Collins:2017oxh,Echevarria:2016scs} up to $n=2$, where they
differ by a factor $2$.  The coefficient $\gamma_K^{(3)}$ has been
recently computed in
Refs.~\cite{Davies:2016jie,Moch:2017uml,Moch:2018wjh}.

\subsection{Logarithmic ordering}
\label{s:pertordering}

In this section, we discuss how to combine in a consistent way the
perturbative ingredients of Eqs.~(\ref{eq:Hexp})-(\ref{eq:andimexp})
for the computation of the cross section in
Eq.~(\ref{eq:crosssection}) (see also
Refs.~\cite{Stewart:2013faa,Ebert:2016gcn}).

As is well known, TMD factorisation provides resummation of large
logarithms of $Q/q_T$ or, equivalently, of $Q/\mu_b$. The resummation
is implemented in the Sudakov form factor $R$ in
Eq.~(\ref{eq:evkernelexp}) whose perturbative expansion reads
\begin{equation}
\label{eq:sudexp}
R = 1 + \sum_{n=1}^\infty
\left( \frac{\alpha_s(Q)}{4\pi} \right)^{n} \sum_{k=1}^{2n} L^k R^{(n,k)} \; ,
\end{equation}
with
\begin{equation}
L = \ln \frac{Q^2}{\mu_b^2} \; .
\end{equation}
Because of the inner sum running up to $2n$, Eq.~(\ref{eq:sudexp})
exposes the double-logarithmic nature of the resummation. This
structure can be traced back to the evolution equations in
Eq.~(\ref{eq:eveqs}) that resum \textit{two different} categories of
logarithms. However, our particular choice of the scales
($\mu_0=\sqrt{\zeta_0}=\mu_b$ and $\mu=\sqrt{\zeta}=Q$) makes the two
categories to coincide, producing up to two logarithms for each power
of $\alpha_s$. Consequently, Eq.~(\ref{eq:sudexp}) must include all
powers of $\alpha_s$ if the scales are such that
$\alpha_s L^2 \gtrsim 1$.

\begin{figure}[tbp]
\centering
\includegraphics[width=0.48\textwidth]{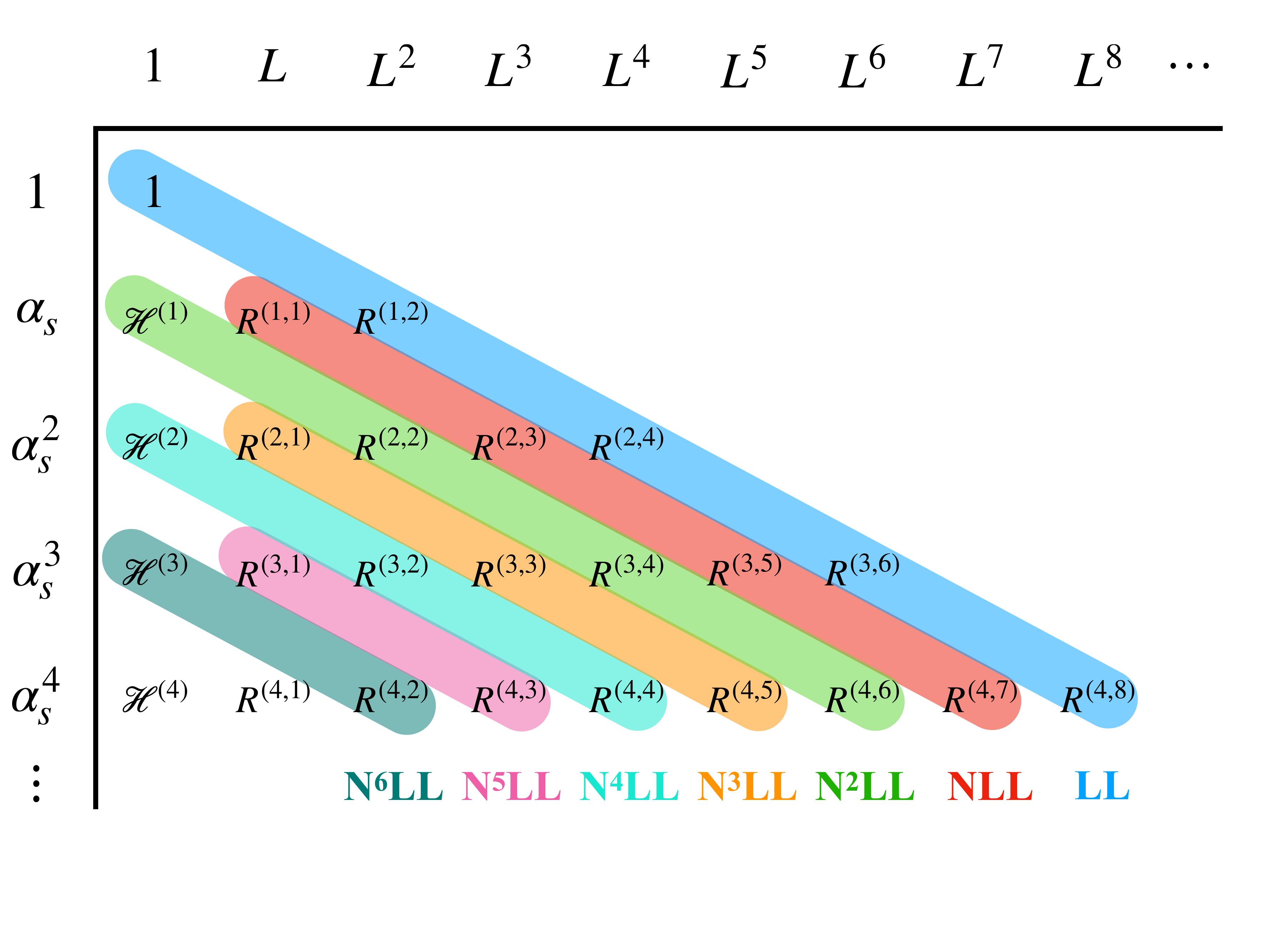}
\includegraphics[width=0.48\textwidth]{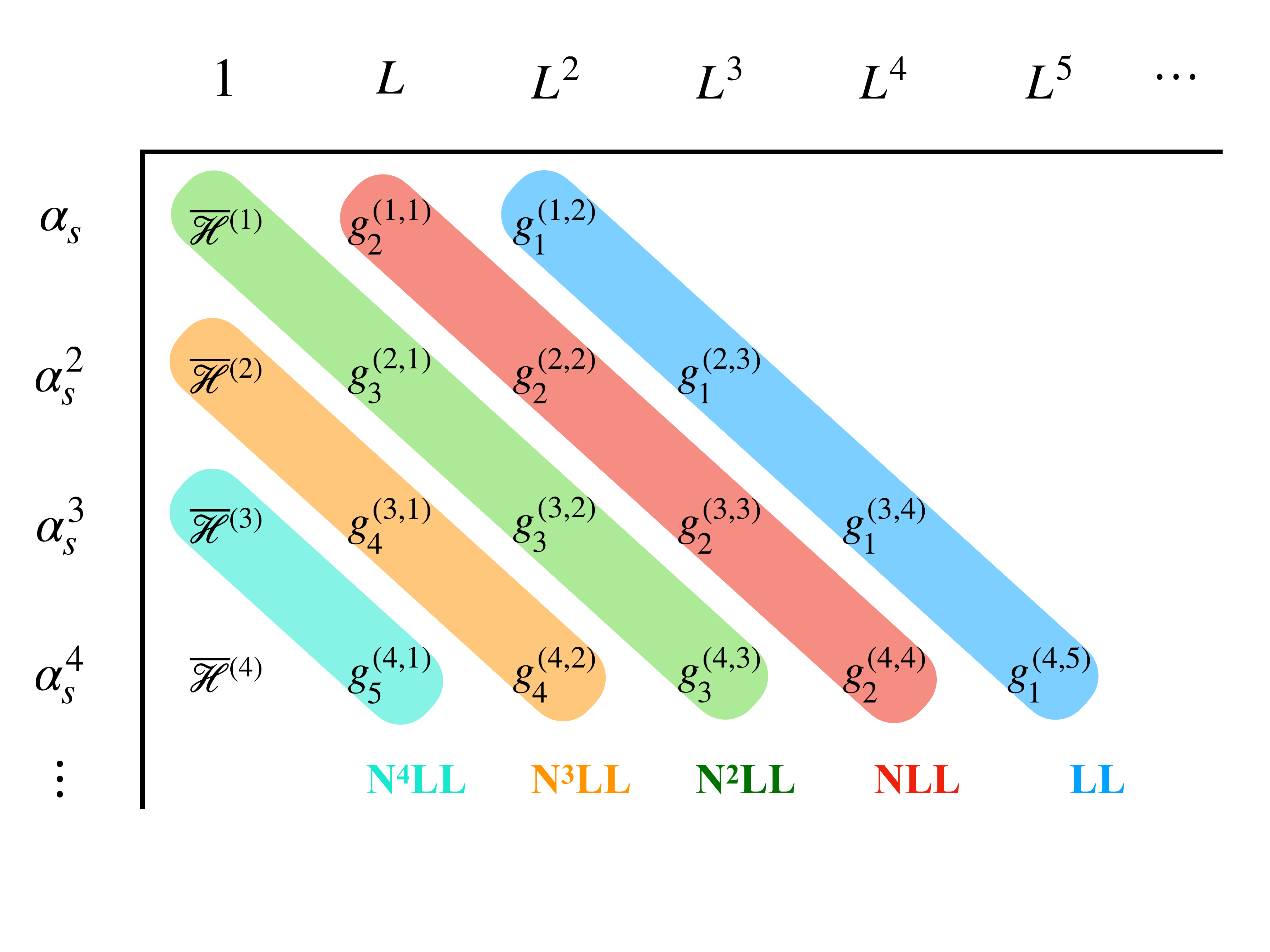}
\caption{Graphical representation of logarithmic countings: in the
  left panel the counting is done at the level of the cross section,
  in the right panel at the level of the logarithm of the cross
  section.}
\label{f:LogTable}
\end{figure}

The expansion~(\ref{eq:sudexp}) can be rearranged to define a
logarithmic ordering as
\begin{equation}
\label{eq:logexp}
R = 1+\sum_{k=0}^{\infty}  R_{{\rm N}^k{\rm LL}} \; ,
\end{equation}
with
\begin{equation}
  R_{{\rm N}^k{\rm LL}} = \sum_{n=1+[k/2]}^\infty \left( \frac{\alpha_s(Q)}{4\pi} \right)^{n} L^{2n-k} R^{(n,2n-k)} \; ,
\end{equation}
where $[k/2]$ is the integer part of $k/2$. According to this
definition, the term $k=0$ in Eq.(\ref{eq:logexp}) gives the
leading-logarithmic (LL) approximation, the term $k=1$ gives the
next-to-leading-logarithmic (NLL) approximation, and so
on. Multiplication of $R_{{\rm N}^k{\rm LL}}$ by a power $p$ of
$\alpha_s$ gives
\begin{equation}
\label{eq:LogAcc}
\left( \frac{\alpha_s(Q)}{4\pi} \right)^p R_{{\rm N}^k{\rm LL}} = \sum_{m=1+[(k+2p)/2]}^\infty
\left( \frac{\alpha_s(Q)}{4\pi} \right)^{m} L^{2m-(k+2p)} R^{(m-p,2m-(k+2p))} \sim R_{{\rm N}^{k+2p}{\rm LL}} \; ,
\end{equation}
where the symbol $\sim$ means that the left- and right-hand sides have
the same logarithmic accuracy. This step is relevant because in the
cross section the Sudakov form factor, Eq.~(\ref{eq:logexp}), can be
multiplied by some power of $\alpha_s$ originating from the hard
factor $H$ and/or the matching functions
$C$. Equation~(\ref{eq:LogAcc}) states that, at the cross section
level, the inclusion of an additional power of $\alpha_s$ in the
perturbative expansion of $H$ and/or $C$ implies a contribution two
orders higher with respect to the leading term in the logarithmic
expansion. For example, at LL and NLL accuracy the functions $H$ and
$C$ can be computed at $\mathcal{O}(1)$, at NNLL and N$^3$LL they need
to include the $\mathcal{O}(\alpha_s)$ corrections, and so on. This
logarithmic counting is illustrated in the left panel of
Fig.~\ref{f:LogTable}: the diagonal bands represent the terms included
in each $R_{{\rm N}^k{\rm LL}}$, with $\mathcal{H}^{(n)}$ the
perturbative coefficients of either $H$ or $C$ or a combination of the
two.

The counting discussed above generally applies to any process whose
amplitude factorises in the appropriate limit, such as DY in the
$q_T\ll Q$ limit (TMD factorisation). However, in the specific case of
DY ({\it i.e.}, inclusive with respect to soft-collinear QCD
radiation) also the phase space for the emission of $n$ real particles
in $\bT$ space factorises (see, \textit{e.g.},
Ref.~\cite{Catani:1996rb}). This feature, along with the factorisation
of the amplitude in the $q_T\ll Q$ limit, allows one to
\textit{exponentiate} soft-collinear emissions such that the Sudakov
form factor can be written in the following general form (see,
\textit{e.g.}, Ref.~\cite{Bozzi:2010xn})\footnote{The factors $1/2$ in
  the argument of the exponential are justified by the fact that each
  of the two TMDs involved in the DY cross section contains an
  evolution factor $R$. In this way, Eq.~(\ref{eq:sudakovCatani})
  matches the literature on $q_T$-resummation where the Sudakov form
  factor is usually defined as the combination of both $R$'s.}
\begin{equation}
  \label{eq:sudakovCatani}
  R = \exp \left[ \frac{1}{2} Lg^{(1)} (\alpha_sL) + \frac{1}{2} g^{(2)} (\alpha_sL) + \frac12 \alpha_s g^{(3)} (\alpha_sL) + \dots \right] \; ,
\end{equation}
where the functions $g^{(i)}$ are such that $g^{(i)}(0)=0$.  As
compared to the general counting in Eq.~(\ref{eq:sudexp}),
exponentiation relates all the terms in Eq.~(\ref{eq:sudexp}) of the
type $\alpha_s^nL^m$ with $n+1 < m \leq 2n$ to the lower-order
terms. In Eq.~(\ref{eq:sudakovCatani}), the logarithmic counting is
performed at the level of the argument of the exponential. In this
context, the terms $Lg^{(1)}$, $g^{(2)}$, $\alpha_sg^{(3)}$, etc.,
resum, respectively, the LL contributions $\alpha_s^n L^{n+1}$, the
NLL contributions $\alpha_s^n L^{n}$, the NNLL contributions
$\alpha_s^n L^{n-1}$, etc.. Contrary to Eq.~(\ref{eq:sudexp}), this
counting is driven by the condition $\alpha_sL\gtrsim 1$.
This extends the validity of the resummed result
(truncated at a given level: NLL, NNLL, etc.)
to larger values of $L$ (smaller values of $q_T/Q$).

The logarithmic counting applied to the argument of the exponential is
equivalent to consider the logarithm of the cross
section~\cite{Bizon:2018foh}. In fact, neglecting for simplicity the
matching functions, we schematically have
\begin{equation}
\label{eq:schematic}
\ln \left( \frac{d\sigma}{dQ dy dq_T} \right) \propto \ln H + L g^{(1)} + g^{(2)} + \alpha_s g^{(3)} + \dots
\end{equation}
The logarithm of $H$ can be expanded as
\begin{equation}
\label{eq:Hlog}
\ln ( 1 + \alpha_s H^{(1)} + \alpha_s^2 H^{(2)} ) = \alpha_s H^{(1)} + \alpha_s^2 \left( H^{(2)} - \frac{H^{(1)2}}{2} \right) + \mathcal{O} ( \alpha_s^3 ) \; .
\end{equation}
The first term $\alpha_s H^{(1)}$ contributes to the tower
$\alpha_s^n L^{n-1}$, that is the NNLL contribution. The second term
$\alpha_s^2 \left( H^{(2)} - H^{(1)2}/2 \right)$ contributes to the
$\alpha_s^n L^{n-2}$ tower, thus to the N$^3$LL contribution. The same
counting applies to the matching functions $C$. The conclusion is that
including $\mathcal{O}(\alpha_s)$ contributions in $H$ and $C$ implies
introducing NNLL corrections, $\mathcal{O}(\alpha_s^2)$ contributions
in $H$ and $C$ contribute to N$^3$LL accuracy, and so on. A graphical
representation of this counting is sketched in the right panel of
Fig.~\ref{f:LogTable}. Again, the bands represent the logarithmic
towers, while $\overline{\mathcal{H}}^{(n)}$ are the appropriate
coefficients of the expansion of either $\ln H$ or $\ln C$ or a
combination. This logarithmic counting has been used in several papers
(see, \textit{e.g.},
Refs.~\cite{Becher:2011xn,Becher:2012yn,Banfi:2016zlc,Bizon:2018foh}). In
this work, we will simply denote this counting with the acronyms NLL,
NNLL, and so on, and for convenience we will refer to it as to
``standard counting''.

A slightly different counting has also been widely used in the
literature (see, \textit{e.g.}, Refs.~\cite{Bozzi:2005wk,
  Catani:2013tia,Stewart:2013faa, Muselli:2017bad, Alioli:2019qzz}).
Expanding the Sudakov form factor~(\ref{eq:sudakovCatani}) and
multiplying it by the expansion of the hard function in
Eq.~(\ref{eq:Hexp}), we obtain for the cross section
\begin{equation}
\label{eq:CataniArgument}
  \frac{d\sigma}{dQ dy dq_T} \propto 1 + L g^{(1)} + g^{(2)} + H^{(1)} \alpha_s L g^{(1)} + \dots \; ,
\end{equation}
where the rightmost term stems from the combination of the first-order
terms $\alpha_s H^{(1)}$ and $L g^{(1)}$ in both expansions.  As it is
clear from the previous discussion, this term has the same form
$\alpha_s^nL^n$ as $g^{(2)}$. Then one can argue that NLL accuracy
requires the inclusion not only of $g^{(2)}$ but also of
$H^{(1)}$~\cite{Bozzi:2005wk}. This argument works to all orders: at
any given logarithmic accuracy, it prescribes to include one more
order in the perturbative expansion of $H$ (and/or $C$) with respect
to the standard counting. We will refer to this counting as the to
``primed counting'', denoting it as NLL$'$, NNLL$'$, and so on. The
apparent contradiction between the standard and primed countings is
resolved by observing that the first term of the perturbative
expansion of $\alpha_s L g^{(1)}$ is proportional to $\alpha_s^2 L^2$.
When considering the general expansion of the cross section given in
Eqs.~(\ref{eq:logexp})-(\ref{eq:LogAcc}), a term proportional to
$\alpha_s^2L^2$ is of the form $\alpha_s^n L^{2n-2}$ and thus belongs
to the NNLL tower. This is formally subleading with respect to the NLL
accuracy determined by the $g^{(2)}$ term in the exponent.

Accurate predictions over a wide range in $q_T$ require matching
resummed calculations (valid at $q_T \ll Q$) to the corresponding
fixed-order calculation (valid at $q_T\lesssim Q$). In this context,
the primed ordering turns out to be more advantageous. Indeed, the
accuracy of a fixed-order calculation is measured in terms of powers
of $\alpha_s$ relative to the leading term. In order to produce a $Z$
boson with large $q_T$, it is necessary to produce (at least) a second
object with large transverse momentum against which the $Z$ boson
recoils, \textit{i.e.}, a jet.  As a consequence, the leading-order
(LO) contribution to the $q_T$ distribution of the $Z$ at fixed order
is $\mathcal{O}(\alpha_s)$. The NLL$'$ prescription correctly
reproduces the small-$q_T$ limit of the LO fixed-order calculation. It
is then possible to realise the matching in an \textit{additive} way
by combining the NLL$'$ resummed calculation with the LO fixed-order
one (NLL$'$ + LO). The procedure can be extended to higher orders:
NNLL$'$ + NLO, N$^3$LL$'$ + NNLO, and so on.  Conversely, in the
standard counting the matching to the LO fixed-order calculation
requires to go further to NNLL accuracy (NNLL + LO), combining in this
way a rather accurate calculation at small $q_T$ with a poorly
accurate calculation at large $q_T$. At higher orders one has N$^3$LL
+ NLO, N$^4$LL + NNLO, and so on. We remark that other forms of
matching can be used to overcome the limitation of the standard
counting~\cite{Bizon:2018foh, Echevarria:2018qyi, Lustermans:2019plv}.

Finally, Tab.~\ref{t:logcountings} summarises the perturbative
ingredients to be used for a consistent computation of the cross
section in Eq.~(\ref{eq:crosssection}) for both the standard and the
primed countings. The numbers in Tab.~\ref{t:logcountings} give the
maximum power of $\alpha_s$ at which the corresponding quantity is to
be computed, while the last column reports the corresponding accuracy
in computing the evolution of the collinear PDFs and of the coupling
$\alpha_s$.\footnote{In the ``unprimed'' counting, $\alpha_s$ is evolved at
  one loop less than the cusp anomalous dimensions for two reasons: first, the
running coupling renormalization group equation resums single logs, therefore the $\beta$ function can be
taken at the same order as the non-cusp anomalous dimension. Secondly, in our
analysis for consistency we take $\alpha_s$ from the LHAPDF grid of the PDF
set we use.}
In this analysis, we have used the PDF sets of the {\tt
  MMHT2014} family~\cite{Harland-Lang:2014zoa} at the appropriate
perturbative order accessed through the {\tt LHAPDF}
interface~\cite{Buckley:2014ana}.
\begin{table}[h!]
\begin{center}
\begin{tabular}{|c|c|c|c|c|}
 \hline
 Accuracy                  &  $H$ and $C$    &  $K$   and  $\gamma_F$
  &  $\gamma_K$  & PDF and $\alpha_s$ evolution              \\
\hline
\hline
LL            & 0      &  -  &     1   &   - \\
 \hline
NLL          & 0     & 1         &    2   & LO\\
 \hline
NLL$'$         & 1     & 1        &   2    & NLO \\
 \hline
NNLL       & 1     & 2         &   3    & NLO \\
 \hline
NNLL$'$      & 2     & 2         &  3    &  NNLO\\
 \hline
N$^3$LL & 2      & 3    &  4   &  NNLO \\
 \hline
\end{tabular}
\caption{Truncation order in the expansions of
  Eqs.~(\ref{eq:Hexp})-(\ref{eq:andimexp}) for the two logarithmic
  countings considered in this paper (see text). The last column
  reports the order used for the evolution of the collinear PDFs and
  $\alpha_s$.}
\label{t:logcountings}
\end{center}
\end{table}

\subsection{Non-perturbative content and its parameterisation}
\label{s:npfunc}

In the previous section, we noticed that in the $\overline{\mbox{MS}}$
scheme the rapidity evolution kernel $K$ and the matching functions
$C$ can be made free of logarithms of the scales by introducing the
natural scale $\mu_b$ defined in Eq.~(\ref{eq:mub}).  Consistently, in
the perturbative expansion of $K$ (see first line of
Eq.~(\ref{eq:andimexp})) and $C$ (see Eq.~(\ref{eq:Cexp})) the strong
coupling $\alpha_s$ must be computed at $\mu_b$. For large values of
$\bT$, $\mu_b$ becomes small such that $\alpha_s(\mu_b)$ may
potentially become very large and eventually diverge when $\mu_b$
reaches the Landau pole at $\Lambda_{\rm QCD}$. As a matter of fact,
the integral in Eq.~(\ref{eq:crosssection}) does require accessing
large values of $\bT$. It is then necessary to regularise this
divergence by introducing a \textit{prescription} that avoids
integrating over the Landau pole. Different possibilities are
available (see, \textit{e.g.}, Refs.~\cite{Catani:1996yz,
  Lustermans:2019plv}). In this paper, we adopt the prescription
originally proposed in Ref.~\cite{Collins:1984kg}: we introduces the
\textit{arbitrary} parameter $\bmax$ that denotes the maximum value of
$\bT$ at which perturbation theory is considered reliable. Hence,
$\bmax$ must be such that
\begin{equation}
\alpha_s \left( \frac{2e^{-\gamma_E}}{\bmax} \right) \ll 1\; .
\end{equation}
Moreover, we also want to prevent $\mu_b$ from becoming much larger
than the hard scale $Q$ ($\mu_b\gg Q$). Despite not strictly mandatory
(especially when considering only small values of $q_T$), this feature
makes it possible to expand the cross section integrated in $q_T$,
with the lowest-order term reproducing the lowest-order collinear
result~\cite{Collins:2016hqq}. To this end, we define
\begin{equation} \label{eq:bmindef}
  \bmin = \frac{2e^{-\gamma_E}}{Q} \; ,
\end{equation}
and introduce a monotonic function $\bstar (\bT)$ with the following
asymptotic behaviours
\begin{equation} \label{eq:bstardef}
\begin{array}{lll}
    \bstar(\bT) \rightarrow \bmin &\mbox{ for } & \bT \rightarrow 0 \;
    , \\ \bstar(\bT) \rightarrow \bmax &\mbox{ for } & \bT \rightarrow
    \infty \; .
\end{array}
\end{equation}
In this analysis, we adopt for $\bstar (\bT)$ the same functional form
chosen in Ref.~\cite{Bacchetta:2017gcc} that guarantees a smooth and
rapid convergence towards the asymptotic
limits: \begin{equation} \label{eq:bstardefPV17} \bstar(\bT) = \bmax
  \left( \frac{1-\exp \left( -\frac{\bT^4}{\bmax^4} \right)}{1-\exp
      \left( -\frac{\bT^4}{\bmin^4} \right)} \right)^{\frac14} \; .
\end{equation}

Now, we simply writes the TMD $\F$ as
\begin{equation}\label{eq:separatation}
\begin{array}{rcl}
\displaystyle  \F(x, \bT ; \mu , \zeta) &=& \displaystyle \left[ \frac{\F(x, \bT ; \mu , \zeta)}{\F (x, \bstar(\bT) ; \mu , \zeta)} \right]
\F(x, \bstar(\bT) ; \mu , \zeta) \\
\\
&\equiv& \displaystyle f_{\rm NP} (x, \bT , \zeta) \F (x, \bstar(\bT) ; \mu , \zeta) \; .
\end{array}
\end{equation}
This separation effectively defines $f_{\rm NP}$.  The advantage is
that, due to the behaviour of $\bstar(\bT)$ for large values of $\bT$,
$\F(x, \bstar(\bT),\mu,\zeta)$ remains in the perturbative region.
The non-perturbative contributions are instead confined into
$f_{\rm NP}$, that has to be determined through a fit to experimental
data. However, using Eq.~(\ref{eq:separatation}), we can work out some
general properties of $f_{\rm NP}$.  First, $f_{\rm NP}$ does not
depend on the renormalisation scale $\mu$. To see this, using
Eqs.~(\ref{eq:solution2}) and~(\ref{eq:evkernelexp}) with
$\mu_0 = \sqrt{\zeta_0}=\mu_b$, we find
\begin{equation}\label{eq:fNP}
\begin{array}{rcl}
  \displaystyle  f_{\rm NP}(x, \bT , \zeta) &=& \displaystyle \frac{\F(x, \bT ; \mu , \zeta)}{\F(x, \bstar(\bT) ; \mu , \zeta)} =
                                                \exp \Bigg\{ K(\mu_b) \ln \frac{\sqrt{\zeta}}{\mu_b} - K(\mu_{b_*}) \ln \frac{\sqrt{\zeta}}{\mu_{b_*}} \\
  \\
                                            &+&\displaystyle \int_{\mu_b}^{\mu_{b_*}} \frac{d\mu'}{\mu'} \left[ \gamma_F (\alpha_s(\mu')) - \gamma_K (\alpha_s(\mu')) \ln
                                                \frac{\sqrt{\zeta}}{\mu'} \right] \Bigg\} \frac{\F (x, \bT ; \mu_b ,\mu_b^2)}{\F (x, \bstar(\bT) ; \mu_{b_*} , \mu_{b_*}^2)} \; ,
\end{array}
\end{equation}
with $\mu_{b_*} \equiv \mu_b(\bstar (\bT))$. The dependence on $\mu$
evidently cancels in the ratio. In addition, for large values of $\bT$
$\mu_{\bstar}$ saturates to some minimal value while $\mu_b$ becomes
increasingly small. As a consequence of this departure between
$\mu_{\bstar}$ and $\mu_b$, as well as between $\sqrt{\zeta}$ and
$\mu_{b}$, the exponential in Eq.~(\ref{eq:fNP}) tends to be
suppressed, and so does $f_{\rm NP}$. Conversely, as $\bT$ becomes
small $\bstar$ approaches $\bmin$. Using the definition in
Eq.~(\ref{eq:bmindef}), it follows that $\mu_{b_*}$ saturates to $Q$
while $\mu_b$ becomes larger and larger. In this limit, we
have~\cite{Collins:2016hqq}
\begin{equation}
f_{\rm NP} \mathop{\longrightarrow}_{\bT \rightarrow 0} 1 + \mathcal{O} \left( \frac{1}{Q^p} \right) \; ,
\end{equation}
where $p$ is some positive number. Since TMD factorisation applies to
leading-power in $q_T/Q$, we can neglect the power suppressed
contribution such that $f_{\rm NP} \to 1$ for $\bT \to 0$. It is
important to stress that the separation between perturbative and
non-perturbative components of a TMD is \textit{arbitrary} and depends
on the particular choice of $\bstar$ (or in general on the
prescription used to regularise the Landau pole). For any given
choice, only the combination in Eq.~(\ref{eq:separatation}) is
meaningful, and it is misleading to refer to $f_{\rm NP}$ as to the
non-perturbative part of TMDs in a universal sense.

Following the requirements discussed above, we parameterise
$f_{\rm NP}$ as
\begin{equation}
\label{eq:fNPparam}
\begin{array}{rcl}
f_{\rm NP} (x, \bT , \zeta) &= &\displaystyle \left[
\frac{1-\lambda}{1 + g_1(x) \frac{\bT^2}{4}} + \lambda \exp \left(- g_{1B}(x) \frac{\bT^2}{4} \right) \right] \\
\\
&\times &\displaystyle \exp \left[ - \left( g_2+ g_{2B} \bT^2 \right) \ln \left( \frac{\zeta}{Q_0^2} \right) \frac{\bT^2}{4} \right] \; ,
\end{array}
\end{equation}
with $Q_0 = 1$~GeV and with the $g_1 (x)$ and $g_{1B} (x)$ functions given by
\begin{equation}
\label{eq:auxfuncs}
\begin{array}{rcl}
g_1(x) &= &\displaystyle  \frac{N_1}{x\sigma} \exp \left[ - \frac{1}{2 \sigma^2} \ln^2 \left( \frac{x}{\alpha} \right) \right] \; , \\
\\
g_{1B}(x) &= &\displaystyle \frac{N_{1B}}{x\sigma_B} \exp \left[ - \frac{1}{2 \sigma_B^2} \ln^2 \left( \frac{x}{\alpha_B} \right) \right] \; .
\end{array}
\end{equation}
There are a total of 9 free parameters
$(\lambda, g_2, g_{2B}, N_1, \sigma, \alpha, N_{1B}, \sigma_B,
\alpha_B)$ to be determined from data.

Apart from the logarithmic dependence on $\zeta$, the functional
form~(\ref{eq:fNPparam}) is motivated by empirical considerations. The
first line parameterises the ``intrinsic'' TMD non-perturbative
contribution and it only depends on $x$ and $\bT$.  The second line
accounts for the non-perturbative correction to the perturbative
evolution. Therefore, it only depends on $\bT$ (on top of the known
dependence on $\zeta$).

The intrinsic contribution is a combination of a $q$-Gaussian (or
Tsallis) distribution (first term) and a standard Gaussian
distribution (second term). The $q$-Gaussian has a larger tail than
the standard Gaussian, meaning that it gives a bigger contribution to
the TMD at small transverse momentum.  We found that this combination
is able to reproduce the behaviour at very small $q_T$ of the
experimental distributions from the lowest to the highest energies
considered in our analysis.

The functions $g_1$ and $g_{1B}$ in Eq.~(\ref{eq:auxfuncs}) are
related to the width of the TMD distribution. Their are expected to
depend on $x$ on the basis of model calculations (see
Ref.~\cite{Burkardt:2015qoa} and references therein) and more
generally from Lorentz invariance constraints on the proton
light-front wave functions (see, \textit{e.g.}, the discussion in
Ref.~\cite{Muller:2014tqa}).  To best describe experimental data, we
found it necessary to have wider TMDs at intermediate $x$.  A
log-normal dependence of $g_1$ and $g_{1B}$ allowed us to properly
describe the datasets differential in the boson rapidity $y$.  In
fact, as we will show below, the $x$ dependence of $f_{\rm NP}$ is
almost entirely determined by the ATLAS datasets, the only ones
differential in $y$. Our present results are quite different from the
ones obtained through fits to semi-inclusive DIS
data~\cite{Bacchetta:2017gcc}.  We expect that the addition of further
datasets from DIS
experiments~\cite{Airapetian:2012ki,Aghasyan:2017ctw} will provide
more sensitivity to the $x$ dependence and possibly lead to different
results.

The non-perturbative components of the TMDs could depend also on
flavour~\cite{Signori:2013mda, Bacchetta:2018lna, Bozzi:2019vnl}.
However, in this work we refrain from including such dependence since
DY data are not very sensitive to it.  We stress that the fact that we
can achieve a good description of data does not exclude the presence
of a flavour dependence, which is actually expected on the basis of
model calculations~\cite{Bacchetta:2008af,Wakamatsu:2009fn,%
  Efremov:2010mt,Bourrely:2010ng,Matevosyan:2011vj,Schweitzer:2012hh},
lattice QCD studies~\cite{Musch:2010ka}, and also if QED corrections
are taken into account~\cite{Bacchetta:2018dcq,Cieri:2018sfk}. Higher
sensitivity to flavour dependence may be provided again by
semi-inclusive DIS data with different targets and final-state hadrons
and possibly by $W$-boson production data~\cite{Lupton:2019mwd}.

Concerning the $\bT$ dependence of the non-perturbative evolution in
the second line of Eq.~(\ref{eq:fNPparam}), we have used a customary
quadratic term~\cite{Davies:1984sp, Meng:1995yn, Meng:1991da,
  Landry:2002ix} with an additional quartic term. The latter
contribution appears to be useful to reproduce the energy evolution
displayed by the data. Other choices of the functional form have been
discussed in, \textit{e.g.},
Refs.~\cite{Aidala:2014hva,Su:2014wpa,Kang:2015msa,Collins:2014jpa}. This
contribution could be also determined using lattice QCD~\cite{Ebert:2018gzl}.

\section{Experimental data}
\label{s:data}

In this section we describe the experimental data included in this
analysis. We considered $q_T$ distributions in DY production from a
variety of datasets. Some of these were already included in the
analysis of Ref.~\cite{Bacchetta:2017gcc}, \textit{i.e.}~data from:
E605~\cite{Moreno:1990sf}, E288~\cite{Ito:1980ev}, CDF
Run~I~\cite{Affolder:1999jh} and Run~II~\cite{Aaltonen:2012fi}, and D0
Run I~\cite{Abbott:1999wk} and Run II~\cite{Abazov:2007ac}. We refer
the reader to Ref.~\cite{Bacchetta:2017gcc} for more details. The new
datasets included in the present analysis are:
\begin{itemize}
\item $Z\rightarrow\mu^+\mu^-$ distribution from D0
  Run~II~\cite{Abazov:2010kn},
\item forward $Z$-production data from the LHCb experiment at
  7~\cite{Abazov:2010kn}, 8~\cite{Aaij:2015gna}, and
  13~\cite{Aaij:2015zlq} TeV,
\item $Z$-production data from the CMS experiment at
  7~\cite{Chatrchyan:2011wt} and 8~\cite{Khachatryan:2016nbe} TeV,
\item $Z$-production data differential in rapidity from the ATLAS
  experiment at 7~\cite{Chatrchyan:2011wt} and 8~\cite{Aad:2015auj}
  TeV,
\item off-peak (low- and high-mass) DY data from the ATLAS experiment
  at 8 TeV~\cite{Aad:2015auj},
\item preliminary $Z$-production data from the STAR experiment at 510
  GeV.\footnote{We thank the STAR Collaboration for providing us with
    the data.}
\end{itemize}
Finally, we originally considered also measurements from the PHENIX
experiment at the center-of-mass energy of 200
GeV~\cite{Aidala:2018ajl}.  However, due to the cut on $q_T/Q$
discussed below, only two data points from this dataset would be
included in the fit. Therefore, we decided to exclude it.

\begin{figure}[t]
  \begin{centering}
    \includegraphics[width=0.8\textwidth]{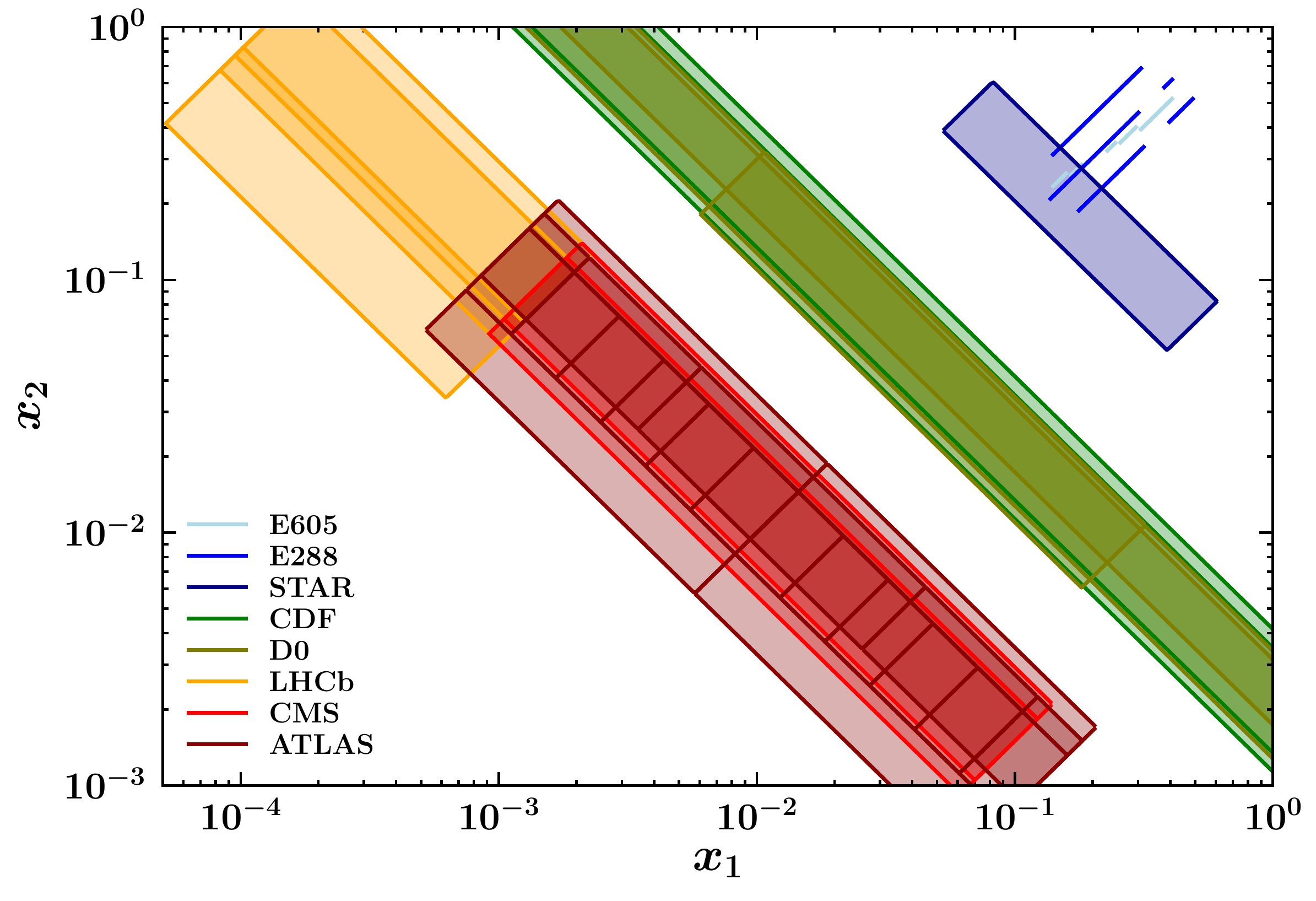}
    \caption{Kinematic coverage on the $x_1$
      vs. $x_2$
      plane of the dataset included in the present
      analysis.\label{fig:KinematicCoverage}}
  \end{centering}
\end{figure}

The breakdown of the entire dataset included in our analysis is
reported in Tab.~\ref{t:data}. For visualisation purposes, in
Fig.~\ref{fig:KinematicCoverage} we show the kinematic coverage of
each datasets in the $x_1$ vs. $x_2$ plane, with
$x_{1,2}=Q e^{\pm y} / \sqrt{s}$. The shaded areas are determined
considering the corresponding ranges in $Q$ and $y$, and the
center-of-mass energy $\sqrt{s}$.\footnote{It should be kept in mind
  that Fig.~\ref{fig:KinematicCoverage} only provides an approximated
  view of the real coverage, strictly true only at tree level. The
  reason is that $x_1$ and $x_2$ are just the lower bounds of
  convolution integrals (see, \textit{e.g.},
  Eq.~(\ref{eq:matching})). Therefore, the effective region of
  sensitivity actually extends between $x_{1,2}$ and 1.}  As expected,
the lower-energy experiments (E605, E288, and STAR) are placed in the
large-$x$ region ($x\gtrsim 0.1$).  Particularly important are the new
(preliminary) STAR measurements that cover a kinematic region that is
scarcely populated. The Tevatron experiments, CDF and D0, cover a
particularly wide kinematic region at intermediate values of
$x$. These experiments (except D0 Run II with muons) provide data
extrapolated over the full range in rapidity $y$, thus extending
across the full available phase space. Finally, the LHC experiments
(LHCb, CMS, and ATLAS) are placed at lower values of $x$. The LHCb
datasets are in a region in which $x_1$ is particularly small and
$x_2$ particularly large: this is due to the fact that the data is
taken in the forward region, $2 < y < 4.5$. The ATLAS datasets are
binned in rapidity and thus are expected to be particularly sensitive
to the $x$ dependence of the TMDs. Indeed, we will show below that the
$x$ dependence of TMDs is mostly constrained by these datasets.
\begin{table}[t]
\footnotesize
\begin{center}
\renewcommand{\tabcolsep}{0.4pc} 
\renewcommand{\arraystretch}{1.2} 
\begin{tabular}{|c|c|c|c|c|c|c|c|}
  \hline
  Experiment & $N_{\rm dat}$ & Observable  &  $\sqrt{s}$ [GeV]& $Q$
                                                         [GeV] &  $y$
                                                                 or
                                                                 $x_F$
  & Lepton cuts & Ref. \\
  \hline
  \hline
  E605 & 50 & $E d^3\sigma/d^3 q$ &  38.8  & 7 - 18  & $x_F=0.1$ & - & \cite{Moreno:1990sf} \\
  \hline
  E288 200 GeV & 30 & $E d^3\sigma/d^3 q$ &  19.4  & 4 - 9  & $y=0.40$ & - & \cite{Ito:1980ev} \\
  \hline
  E288 300 GeV & 39 & $E d^3\sigma/d^3 q$ &  23.8  & 4 - 12  & $y=0.21$ & - & \cite{Ito:1980ev} \\
  \hline
  E288 400 GeV & 61 & $E d^3\sigma/d^3 q$ &  27.4  & 5 - 14  & $y=0.03$ & - & \cite{Ito:1980ev} \\
  \hline
  STAR 510 & 7 & $d\sigma/dq_T$ & 510  & 73 - 114  &
                                                     $|y|<1$
  & \makecell{$p_{T\ell} > 25$~GeV\\ $|\eta_\ell|<1$} & - \\
  \hline
  CDF Run I & 25 & $d\sigma/dq_T$ & 1800 & 66 - 116  & Inclusive & - & \cite{Affolder:1999jh} \\
  \hline
  CDF Run II & 26 & $d\sigma/dq_T$ & 1960 & 66 - 116  &  Inclusive & - & \cite{Aaltonen:2012fi} \\
  \hline
  D0 Run I & 12 & $d\sigma/dq_T$ & 1800 & 75 - 105  &  Inclusive & - & \cite{Abbott:1999wk} \\
  \hline
  D0 Run II & 5 & $(1/\sigma)d\sigma/dq_T$ & 1960 & 70 - 110  & Inclusive & - & \cite{Abazov:2007ac} \\
  \hline
  D0 Run II $(\mu)$ & 3 & $(1/\sigma)d\sigma/dq_T$ & 1960 & 65 - 115  & $|y|<1.7$ & \makecell{$p_{T\ell} > 15$~GeV\\$|\eta_\ell|<1.7$} & \cite{Abazov:2010kn} \\
  \hline
  LHCb 7 TeV & 7 & $d\sigma/dq_T$ & 7000 & 60 - 120  & $2<y<4.5$ & \makecell{$p_{T\ell} > 20$ GeV\\$2<\eta_\ell<4.5$} & \cite{Aaij:2015gna} \\
  \hline
  LHCb 8 TeV & 7 & $d\sigma/dq_T$ & 8000 & 60 - 120  & $2<y<4.5$ & \makecell{$p_{T\ell} > 20$ GeV\\$2<\eta_\ell<4.5$} & \cite{Aaij:2015zlq} \\
  \hline
  LHCb 13 TeV & 7 &  $d\sigma/dq_T$ & 13000 & 60 - 120  &
                                                          $2<y<4.5$ & \makecell{$p_{T\ell} > 20$ GeV\\$2<\eta_\ell<4.5$} & \cite{Aaij:2016mgv} \\
  \hline
  CMS 7 TeV & 4 & $(1/\sigma)d\sigma/dq_T$  & 7000 & 60 - 120  & $|y|<2.1$ & \makecell{$p_{T\ell} > 20$ GeV\\$|\eta_\ell|<2.1$} & \cite{Chatrchyan:2011wt} \\
  \hline
  CMS 8 TeV & 4 & $(1/\sigma)d\sigma/dq_T$ & 8000 & 60 - 120  & $|y|<2.1$ & \makecell{$p_{T\ell} > 15$ GeV\\$|\eta_\ell|<2.1$} & \cite{Khachatryan:2016nbe} \\
  \hline
  ATLAS 7 TeV & \makecell{6\\6\\6}
             &$(1/\sigma)d\sigma/dq_T$ & 7000 & 66 - 116
                                                       &
                                                         \makecell{$|y|<1$ \\ $1<|y|<2$ \\ $2<|y|<2.4$}  & \makecell{$p_{T\ell} > 20$~GeV\\$|\eta_\ell|<2.4$} & \cite{Aad:2014xaa} \\
  \hline
  \makecell{ATLAS 8 TeV \\ on-peak} & \makecell{6\\6\\6\\6\\6\\6} &
                                                                    $(1/\sigma)d\sigma/dq_T$
                      & 8000 & 66 - 116  & \makecell{$|y|<0.4$ \\ $0.4<|y|<0.8$ \\ $0.8<|y|<1.2$\\$1.2<|y|<1.6$\\$1.6<|y|<2$\\$2<|y|<2.4$} & \makecell{$p_{T\ell} > 20$~GeV\\$|\eta_\ell|<2.4$} & \cite{Aad:2015auj} \\
  \hline
  \makecell{ATLAS 8 TeV \\ off-peak} & \makecell{4 \\ 8} &
                                                           $(1/\sigma)d\sigma/dq_T$
                      & 8000 & \makecell{46 - 66 \\ 116 - 150} & $|y|<2.4$ & \makecell{$p_{T\ell} > 20$ GeV\\$|\eta_\ell|<2.4$} & \cite{Aad:2015auj} \\
  \hline
  \hline
  Total & 353 &-&-&-&-&-&-\\
  \hline
\end{tabular}
\caption{Breakdown of the datasets included in this analysis. For each
  dataset, the table includes information on: the number of data
  points ($N_{\rm dat}$) passing the nominal cut on $q_T/Q$, the
  observable delivered, the center of mass energy $\sqrt{s}$, the
  range(s) in invariant mass $Q$, the angular variable (either $y$ or
  $x_F$), possible cuts on the single final-state leptons, and the
  public reference (when available). The total number of data points
  amounts to 353. Note that for E605 and E288 400 GeV we have excluded
  the bin in $Q$ containing the $\Upsilon$ resonance
  ($Q\simeq 9.5$~GeV).}
\label{t:data}
\end{center}
\end{table}

Since our analysis is based on the TMD factorisation formula in
Eq.~(\ref{eq:crosssection}), only data at small $q_T$ can possibly be
described. Hence, we impose a cut to exclude measurements with large
$q_T$ by requiring $q_T / Q< 0.2$. Since the measurements are
delivered in transverse-momentum bins
$[q_{T,\rm min}{:}\;q_{T,\rm max}]$ integrated over some range in
invariant mass $[Q_{\rm min}{:}\;Q_{\rm max}]$, the cut is
conservatively imposed on the ratio $q_{T,\rm max}/Q_{\rm min}$. The
second column in Tab.~\ref{t:data} reports the number of data points
($N_{\rm dat}$) for each dataset that pass this cut: the total number
of points included in our analysis is 353.

An important feature of all the new datasets listed above is that the
cross sections are given within a certain fiducial region. In
particular, kinematic cuts on transverse momentum $p_{T\ell}$ and
pseudo-rapidity $\eta_{\ell}$ of the final-state leptons are
enforced. The values of the cuts are reported in the next-to-last
column of Tab.~\ref{t:data}.  Our predictions are corrected by means
of the phase-space reduction factor $\mathcal{P}$ introduced in
Eq.~(\ref{eq:crosssection}), which takes into account these cuts.
Details concerning the calculation of $\mathcal{P}$ are given in
Appendix~\ref{a:LeptonCuts}.

As evident from the ``Observable'' column of Tab.~\ref{t:data},
experimental cross sections are released in different forms. In
addition, some of them are normalised to the total (fiducial) cross
section while others are not. In our analysis, we expressed all the
absolute cross sections in terms of the observable given in
Eq.~(\ref{eq:crosssection}) (details on the transformations between
different observables can be found in
Ref.~\cite{Bacchetta:2017gcc}). When necessary, the total cross
section $\sigma$ required to normalise the differential cross sections
is computed using {\tt DYNNLO}~\cite{Catani:2007vq,Catani:2009sm} with
the {\tt MMHT2014} collinear PDF sets~\cite{Harland-Lang:2014zoa},
taking into account the selection cuts and consistently with the
perturbative order of the differential cross section. More precisely,
the total cross section is computed at LO for NLL accuracy, at
NLO for NLL' and NNLL, and at NNLO for NNLL' and N$^3$LL. The values of the total cross
sections at different orders are reported in
Tab.~\ref{t:totalcrosssections}. We stress that in this analysis no
additional normalisations have been applied, with the consequence that
both the shape and the normalisation of the experimental distributions
have an impact on the fit.
\begin{table}[t]
\footnotesize
\begin{center}
\renewcommand{\tabcolsep}{0.4pc} 
\renewcommand{\arraystretch}{1.2} 
\begin{tabular}{|c|c|c|c|}
  \hline
  Experiment & LO [pb] & NLO [pb]  &  NNLO [pb]  \\
  \hline
  \hline
  D0 Run II & 170.332 & 242.077 & 253.573 \\
  \hline
  D0 Run II $(\mu)$ & 100.765  & 119.002 & 124.675 \\
  \hline
  CMS 7 TeV & 291.977  & 384.569 & 398.853 \\
  \hline
  CMS 8 TeV & 340.132  & 456.337 & 473.411 \\
  \hline
  \makecell{ATLAS 7 TeV} \makecell{$|y|<1$ \\ $1<|y|<2$ \\
  $2<|y|<2.4$} & \makecell{196.457 \\135.511 \\12.568 } &
                                                                \makecell{251.296 \\181.267 \\17.091 } & \makecell{253.781 \\181.466 \\17.104 } \\
  \hline
  \makecell{ ATLAS 8 TeV\\on-peak} \makecell{$|y|<0.4$ \\ $0.4<|y|<0.8$ \\
  $0.8<|y|<1.2$\\$1.2<|y|<1.6$\\$1.6<|y|<2$\\$2<|y|<2.4$} &
                                                            \makecell{89.531 \\89.120 \\85.499 \\69.018 \\43.597 \\14.398
  } & \makecell{113.650 \\112.853 \\109.800 \\91.884
  \\59.114 \\19.574 } &
                              \makecell{116.766 \\115.738 \\112.457 \\95.187 \\62.127 \\20.937 }\\
  \hline
  \makecell{ ATLAS 8 TeV\\off-peak} \makecell{$46\mbox{ GeV}< Q < 66\mbox{
  GeV}$\\$116\mbox{ GeV}< Q < 150\mbox{ GeV}$}& \makecell{15.199
                                                \\3.805 } &
                                                              \makecell{14.449 \\5.317 } & \makecell{14.368 \\5.521 }\\
  \hline
\end{tabular}
\caption{Total (fiducial) cross sections computed with {\tt
    DYNNLO}~\cite{Catani:2007vq,Catani:2009sm} using the central
  member of the {\tt MMHT2014} collinear PDF
  sets~\cite{Harland-Lang:2014zoa} and required for the computation of
  the normalised differential cross sections at the different
  perturbative orders.\label{t:totalcrosssections}}
\end{center}
\end{table}

Most of the considered experimental datasets are released with a set
of \textit{uncorrelated} and \textit{correlated} uncertainties. As
already pointed out in Ref.~\cite{Bertone:2019nxa}, a proper treatment
of the experimental uncertainties is crucial to achieve a reliable
extraction of TMDs. In other words, the $\chi^2$, which quantifies the
agreement between data and predictions and is minimised during the
fit, has to be computed taking into account the nature of the various
uncertainties. Particular care has to be taken with the (correlated)
\textit{normalisation} uncertainties. As is well known, an
inappropriate description of normalisation uncertainties may lead to
underestimate the predictions: that is the so-called D'Agostini
bias~\cite{DAgostini:1993arp,DAgostini:2003syq}. Different
prescriptions have been devised to avoid this
problem~\cite{Ball:2012wy}: in this analysis we adopt the so-called
iterative $t_0$-prescription~\cite{Ball:2009qv}.

In the presence of correlated uncertainties, the $\chi^2$ can be
split as~\cite{Ball:2012wy}
\begin{equation}\label{eq:chi2sep}
\chi^2 = \chi_D^2 + \chi_\lambda^2\,,
\end{equation}
where $\chi_D^2$ has an uncorrelated structure (diagonal) while
$\chi_\lambda^2$ is a penalty term related to the presence of
correlations (see, \textit{e.g.}, Appendix~B of
Ref.~\cite{Bertone:2019nxa}). For the computation of
$\chi_D^2$, theoretical predictions are properly shifted to take into
account the effect of the correlated uncertainties. In fact, shifted
predictions are a better proxy for visual comparisons to experimental
data. Therefore, in the following it is understood that all plots will display
shifted predictions.

A further important aspect is the use of collinear PDFs. In order to
extract $f_{\rm NP}$ defined in Eq.~(\ref{eq:separatation}), it is
necessary to assume a given set of collinear PDFs ({\tt MMHT2014} in
our case). PDF uncertainties reflect the experimental uncertainty of
the dataset used for their extraction. It is therefore natural to
attribute an experimental nature to this uncertainty and include it in
the calculation of the $\chi^2$. To do so, we computed the PDF errors
as relative to the central value\footnote{The advantage of computing
  relative uncertainties is that of minimising the dependence on the
  non-perturbative function $f_{\rm NP}$ assumed for the computation
  of both the central PDF set and the error members. We also notice
  that the calculation of such uncertainties does include the PDF
  uncertainty on the total cross sections when normalised
  distributions are considered.}  and included them in the
experimental covariance matrix as uncorrelated uncertainties. The
propagation of the resulting experimental uncertainty into the fitted
TMDs is achieved through Monte Carlo sampling. Specifically, we
generate $N_{\rm rep}$ ($\gtrsim 200$) replicas of the original
dataset taking into account all the uncertainties and then perform a
fit on each single replica. The resulting ensemble of distributions
can be used to compute central values and uncertainties as averages
and correlations, respectively.

A final remark concerns the integration over the final-state phase
space. The basic quantity to be compared to data is
\begin{equation}\label{eq:phasespaceintegral}
  \frac{d\sigma}{dq_T} = \frac{1}{q_{T,\rm max}-q_{T,\rm
      min}}\int_{y_{\rm min}}^{y_{\rm max}} dy \int_{Q_{\rm
      min}}^{Q_{\rm max}} dQ \int_{q_{T,\rm min}}^{q_{T,\rm max}}
  dq_T\left[\frac{d\sigma}{dQ dy dq_T}\right]\,,
\end{equation}
where the ranges $[y_{\rm min}{:}\;y_{\rm max}]$,
$[Q_{\rm min}{:}\;Q_{\rm max}]$, and
$[q_{T,\rm min}{:}\;q_{T,\rm max}]$ define the phase-space integration
region and the integrand is given in Eq.~(\ref{eq:crosssection}). In
order to speed up the numerical computation of the theoretical
predictions, the integration over the bins in $q_T$ and $Q$ is often
performed approximating the $q_T$-bin integral with its central value
and using the narrow-width approximation for the integral over $Q$
around the $Z$ peak. We stress that in this analysis the integrals in
Eq.~(\ref{eq:phasespaceintegral}) are computed \textit{exactly}. While
the integrals over $y$ and $Q$ do need to be computed numerically, the
integral over $q_T$ can be performed (semi)analytically exploiting a
property of the Bessel functions $J_n$ (see
Appendix~\ref{a:qTintegration}). This greatly reduces the amount of
numerical computations.

\section{Results}
\label{s:results}

In this section, we present the results of our extraction of
unpolarised TMDs from a comprehensive set of DY data (see
Sec.~\ref{s:data}). In Sec.~\ref{ss:fitquality}, we present the
quality of the fit at N$^3$LL, the best accuracy we can presently
reach. In Sec.~\ref{ss:tmddists} we discuss the TMDs extracted from
the nominal fit. In Sec.~\ref{ss:pertconv}, we discuss the convergence
of the perturbative corrections. In Sec.~\ref{ss:reddata}, we focus on
the $x$ dependence of the TMDs and we argue that it is mostly
constrained by the $y$-differential ATLAS cross sections. Finally, in
Sec.~\ref{ss:cutscan}, we assess the range of validity of TMD
factorisation by considering the fit quality as a function of the cut
on $q_T/Q$.

\subsection{Fit quality}\label{ss:fitquality}

In this section, we discuss the quality of the reference fit at
N$^3$LL with cut $q_T/Q<0.2$. In order to quantify this quality, the
$\chi^2$s are evaluated using the mean of the TMDs extracted from the
Monte Carlo replicas of the data. Denoting the Monte Carlo ensemble of
TMDs with $\{\F^{q,[k]}\}$, $k=1,\dots,N_{\rm rep}$ ($N_{\rm rep}$
being the number of replicas), the \textit{mean} is defined as
\begin{equation}\label{eq:meanreplica}
  \F^q(x,\bT;\mu,\zeta) = \frac{1}{N_{\rm rep}}\sum_{k=1}^{N_{\rm rep}}\F^{q,[k]}(x,\bT;\mu,\zeta)\,.
\end{equation}
The mean value provides a \textit{democratic} representative of the
ensemble. Other choices are possible, such as the median or the mode
of the ensemble. In fact, only the full ensemble of replicas carries
the full statistical information. However, the reason for using
Eq.~(\ref{eq:meanreplica}) is that quantifying the goodness of our fit
becomes easier, as it will be clear in the following.
\begin{table}[htbp!]
\footnotesize
\centering
\renewcommand{\tabcolsep}{0.4pc} 
\begin{tabular}{lcccc}
  \hline
  Experiment            & & $\chi^2_D/N_{\rm dat}$   &
                                                       $\chi^2_{\lambda}/N_{\rm dat}$ &  $\chi^2/N_{\rm dat}$ \\
  \hline
  \hline
  E605                   &\makecell{   7~GeV $< Q <$ 8~GeV
                                    \\ 8~GeV $< Q <$ 9~GeV
                                    \\ 10.5~GeV $< Q <$ 11.5~GeV
                                    \\ 11.5~GeV $< Q <$ 13.5~GeV
                                    \\ 13.5~GeV $< Q <$ 18~GeV }&
\makecell{
0.419\\
0.995\\
0.191\\
0.491\\
0.491}&
\makecell{
0.068\\
0.034\\
0.137\\
0.284\\
0.385}&
\makecell{
0.487\\
1.029\\
0.328\\
0.775\\
0.877}\\\hline
  E288 200 GeV             &\makecell{   4~GeV $< Q <$ 5~GeV
                                    \\ 5~GeV $< Q <$ 6~GeV
                                    \\ 6~GeV $< Q <$ 7~GeV
                                    \\ 7~GeV $< Q <$ 8~GeV
                                    \\ 8~GeV $< Q <$ 9~GeV  }&
\makecell{
0.213\\
0.673\\
0.133\\
0.254\\
0.652}&
\makecell{
0.649\\
0.292\\
0.141\\
0.014\\
0.024}&
\makecell{
0.862\\
0.965\\
0.275\\
0.268\\
0.676}\\ \hline
  E288 300 GeV           &\makecell{   4~GeV $< Q <$ 5~GeV
                                    \\ 5~GeV $< Q <$ 6~GeV
                                    \\ 6~GeV $< Q <$ 7~GeV
                                    \\ 7~GeV $< Q <$ 8~GeV
                                    \\ 8~GeV $< Q <$ 9~GeV
                                    \\ 11~GeV $< Q <$ 12~GeV }&
\makecell{
0.231\\
0.502\\
0.315\\
0.056\\
0.530\\
1.047}&
\makecell{
0.555\\
0.204\\
0.063\\
0.030\\
0.017\\
0.167}&
\makecell{
0.785\\
0.706\\
0.378\\
0.086\\
0.547\\
1.215}\\ \hline
  E288 400 GeV            &\makecell{   5~GeV $< Q <$ 6~GeV
                                    \\ 6~GeV $< Q <$ 7~GeV
                                    \\ 7~GeV $< Q <$ 8~GeV
                                    \\ 8~GeV $< Q <$ 9~GeV
                                    \\ 11~GeV $< Q <$ 12~GeV
                                    \\ 12~GeV $< Q <$ 13~GeV
                                    \\ 13~GeV $< Q <$ 14~GeV }&
\makecell{
0.312\\
0.100\\
0.018\\
0.437\\
0.637\\
0.788\\
1.064}&
\makecell{
0.065\\
0.005\\
0.011\\
0.039\\
0.036\\
0.028\\
0.044}&
\makecell{
0.377\\
0.105\\
0.029\\
0.477\\
0.673\\
0.816\\
1.107}\\ \hline
 \hline
  STAR                         & &  0.782&0.054&0.836\\ \hline\hline
  CDF Run I                 &  &  0.480&0.058&0.538\\ \hline
  CDF Run II                &  &  0.959&0.001&0.959\\ \hline
  D0 Run I                   &  &  0.711&0.043&0.753\\ \hline
  D0 Run II                  &  &  1.325&0.612&1.937\\ \hline
  D0 Run II $(\mu)$    &  &  3.196&0.023&3.218\\ \hline\hline
  LHCb 7 TeV              &  &  1.069&0.194&1.263\\ \hline
  LHCb 8 TeV              &  &  0.460&0.075&0.535\\ \hline
  LHCb 13 TeV            &  &  0.735&0.020&0.755\\ \hline 
  CMS 7 TeV                &  & 2.131&0.000&2.131\\ \hline
  CMS 8 TeV                & &  1.405&0.007&1.412\\ \hline
  ATLAS 7 TeV         & \makecell{   $0<|y|<1$
                                    \\ $1<|y|<2$
                                    \\ $2<|y|<2.4$  }&
\makecell{
2.581\\
4.333\\
3.561}&
\makecell{
0.028\\
1.032\\
0.378}&
\makecell{
2.609\\
5.365\\
3.939}\\ \hline
  \makecell{ ATLAS 8 TeV\\on-peak}        & \makecell{   $0<|y|<0.4$
                                    \\ $0.4<|y|<0.8$
                                    \\ $0.8<|y|<1.2$
                                    \\ $1.2<|y|<1.6$
                                    \\ $1.6<|y|<2$
                                    \\ $2<|y|<2.4$  }&
\makecell{
1.924\\
2.342\\
0.917\\
0.912\\
0.721\\
0.932}&
\makecell{
0.337\\
0.247\\
0.061\\
0.095\\
0.092\\
0.348}&
\makecell{
2.262\\
2.590\\
0.978\\
1.006\\
0.814\\
1.280}\\ \hline
  \makecell{ ATLAS 8 TeV\\off-peak}         & \makecell{   46~GeV $< Q <$ 66~GeV
                                    \\ 116~GeV $< Q <$ 150~GeV }&
\makecell{
2.138\\
0.501}&
\makecell{
0.745\\
0.003}&
\makecell{
2.883\\
0.504}\\ \hline
  \hline
  \textbf{Global}                                &  &  \textbf{0.88} & \textbf{0.14} & \textbf{1.02} \\
  \hline
\end{tabular}
\vspace{-0.2cm}
\caption{The $\chi^2/N_{\rm dat}$ using the mean
replica in Eq.~(\ref{eq:meanreplica}).
$N_{\rm dat}$
in each case is listed in Tab.~\ref{t:data}. The
uncorrelated ($\chi_D^2$) and correlated ($\chi_\lambda^2$)
  contributions and their sum $\chi^2$ are shown (see
  Eq.~(\ref{eq:chi2sep})). \label{t:chi2mean}}
\centering
\end{table}

 Tab.~\ref{t:chi2mean} reports the breakdown of
the $\chi^2$s normalised to the number of data points, $N_{\rm dat}$,
for each dataset. The uncorrelated ($\chi_D^2$) and the correlated
($\chi_\lambda^2$) contributions to the total $\chi^2$ (see
Eq.~(\ref{eq:chi2sep})) are also reported. The global $\chi^2$ is
shown at the bottom of the table.

The value of the global $\chi^2$ is very close to one (1.02),
indicating that the fit is able to describe measurements over a wide
energy range, from the low-energy fixed-target datasets to the LHC
ones. It is important to stress that a substantial contribution to the
global $\chi^2$ is given by the correlated penalty term,
$\chi_\lambda^2/N_{\rm dat} = 0.14$. This highlights the importance of
a correct treatment of the correlated uncertainties. More
specifically, the systematic shifts induced by correlations are often
large, indicating that the fit does need to adjust the predictions
within the experimentally correlated ranges.

Concerning the single experiments, we observe that the low-energy data
(E605, E288, and STAR) have generally lower $\chi^2$s than the
Tevatron (CDF and D0) and LHC (LHCb, CMS, and ATLAS) high-energy
data. This is mostly due to the fact that the experimental
uncertainties of the former are typically larger than the latter. In
particular, the low-energy data are affected by large normalisation
(correlated) uncertainties. Consequently, the relative importance of
the correlated contribution $\chi_\lambda^2$ to the total $\chi^2$ is
generally larger for the low-energy datasets than for the high-energy
ones.

It is interesting to comment on the quality of the fit to the new
datasets from RHIC and the LHC that were not included in the analysis
of Ref.~\cite{Bacchetta:2017gcc} (see Sec.~\ref{s:data}). The
preliminary measurements from STAR have a $\chi^2$ equal to
$0.836$. This is particularly encouraging because, as clear from
Fig.~\ref{fig:KinematicCoverage}, this dataset covers a scarcely
populated kinematic region and shows no tension with other data. Also
the LHC datasets extend the kinematic coverage of the DY data
considered in Ref.~\cite{Bacchetta:2017gcc}. These measurements are
particularly precise and thus very effective in constraining TMDs.  We
observe that the LHCb datasets are very nicely described with
$\chi^2$s that never exceed 1.3. The CMS data, despite having slightly
larger $\chi^2$, are also well described. The two CMS datasets provide
only eight points in total and thus their impact on the fit is
modest. The ATLAS datasets, amongst the LHC ones, are by far the most
abundant.
We observe that the ATLAS 8 TeV
datasets are well described, except for the first two low-rapidity
bins. The 7 TeV ones present larger values of $\chi^2$, above 2.
Given the extremely
high precision of these datasets, even
small effects (e.g., power corrections) could give a significant contribution
to $\chi^2$ in these conditions. 
We consider it already a success to obtain a
value of $\chi^2$ for these datasets
that does not affet too much the global $\chi^2$. 
We note that a key feature of these datasets (except the off-peak ones)
is that they are differential in the vector-boson rapidity $y$. As we
will see in Sec.~\ref{ss:reddata}, the $x$ dependence of $f_{\rm NP}$ plays a
crucial role in improving the $\chi^2$.

\begin{figure}[htb!]
  \begin{centering}
    \includegraphics[width=1\textwidth]{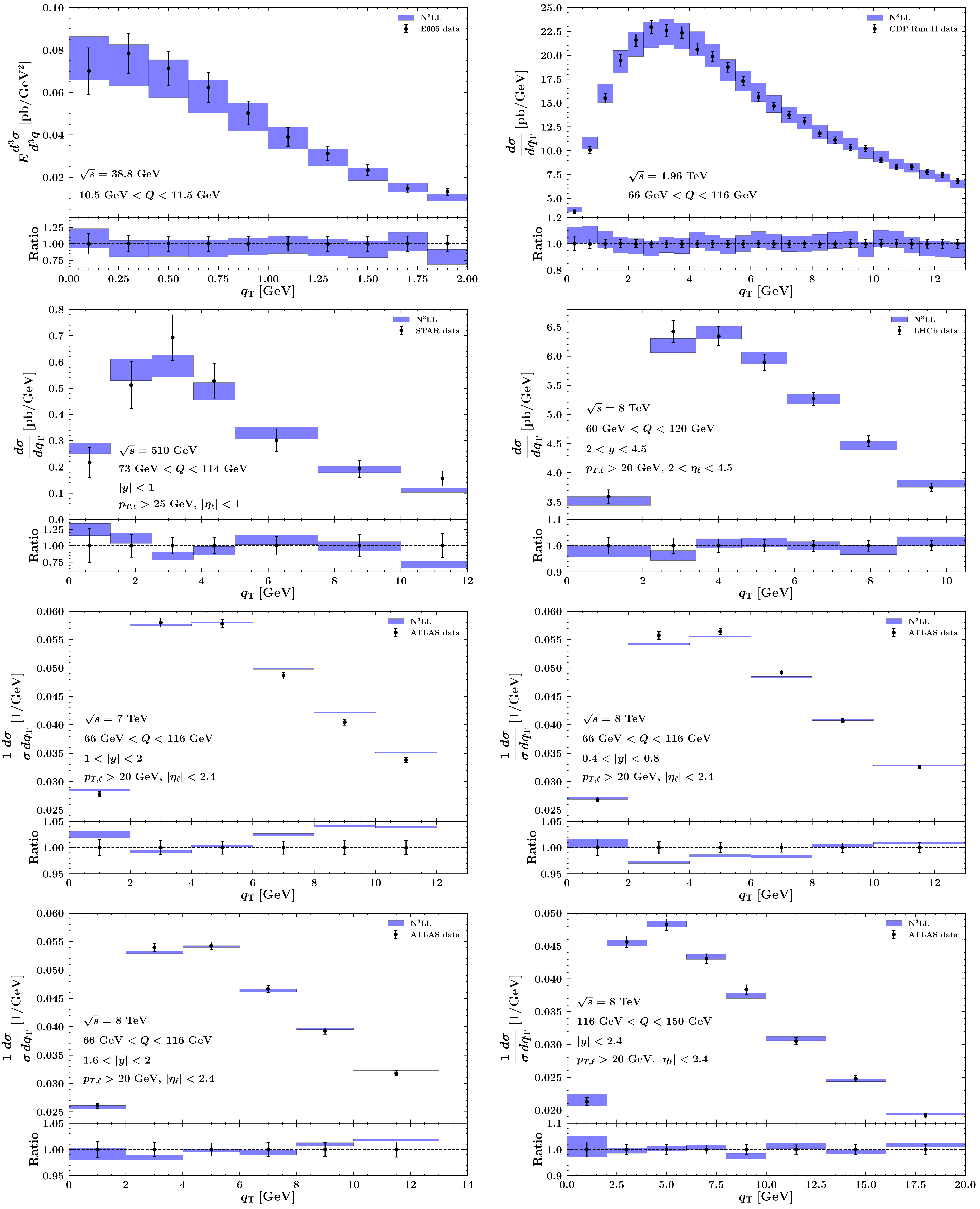}
    \caption{Comparison between experimental data and theoretical
      predictions obtained at N$^3$LL accuracy for a representative
      subset of the datasets included in this analysis. The upper
      panel of each plot displays the absolute $q_{T}$ distributions,
      while the lower panel displays the same distributions normalised
      to the experimental central values. The blue bands represent the
      1-$\sigma$ uncertainty of the theoretical predictions.
      \label{fig:DataTheoryComparison}}
  \end{centering}
\end{figure}
In order to provide a visual assessment of the fit quality,
Fig.~\ref{fig:DataTheoryComparison} displays the data/theory
comparison for a representative selection of datasets. We remind the
reader that in each plot theoretical predictions are appropriately
shifted to account for correlated
uncertainties~\cite{Bertone:2019nxa}, while the experimental error
bars are given by the sum in quadrature of the uncorrelated
uncertainties. The upper panel of each plot shows the absolute $q_T$
distribution, while the lower panel shows the ratio to data. The plots
in the upper row of Fig.~\ref{fig:DataTheoryComparison} refer to one
invariant-mass bin of E605 and CDF Run II already considered in
Ref.~\cite{Bacchetta:2017gcc}. The remaining plots refer to some of
the new datasets, namely STAR, LHCb 8 TeV, ATLAS 8 TeV on-peak at
$1.6<|y|<2$, and ATLAS 8 TeV off-peak at 116 GeV $< Q < $ 150 GeV. As
expected, there is a very good agreement between data and theory, for
both the old and the new datasets. Finally, it is interesting to
observe that the uncertainties of the upper and middle rows of
Fig.~\ref{fig:DataTheoryComparison} are larger than those in the
two lower rows. This is due to the fact that the ATLAS distributions are
normalised to the total cross section leading to a cancellation of
some uncertainties, such as those due to luminosity and collinear
PDFs.


\subsection{TMD distributions}\label{ss:tmddists}

We discuss now the TMD distributions extracted from our reference
N$^3$LL fit. We stress once again that only the combination in the
r.h.s. of Eq.~(\ref{eq:separatation}) is meaningful.

In order to assess the sensitivity of the experimental dataset to
$f_{\rm NP}$, it is interesting to look at the values of the free
parameters obtained from the fit. In Tab.~\ref{t:fittedparameters} the
average of each parameter over the Monte Carlo replicas, along with
the respective standard deviation, is reported. All parameters are
well constrained.\footnote{We stress that the parameters reported in
  Tab.~\ref{t:fittedparameters} are not meant to be used in the
  parameterisation in Eqs.~(\ref{eq:fNPparam})-(\ref{eq:auxfuncs}) as
  they are not a direct result of any of our fits.} It is interesting
to observe that the parameter $\lambda$, that measures the relative
weight of Gaussian and $q$-Gaussian in Eq.~(\ref{eq:fNPparam}), is
close to 0.5 indicating that these contributions weigh approximately
the same. Concerning the values of the parameters $g_2$ and $g_{2B}$
associated to the non-perturbative contribution to TMD evolution, we
find that the coefficient $g_{2B}$ of the quartic term is small but
significantly different from zero. This seems to suggest that
higher-power corrections to the commonly assumed quadratic term $g_2$
may be required by the data.

Further insight concerning the appropriateness of the functional form
in Eqs.~(\ref{eq:fNPparam})-(\ref{eq:auxfuncs}) can be gathered by
looking at the statistical correlations between
parameters. In the right panel of Tab.~\ref{t:fittedparameters},
we show a graphical
representation of the correlation matrix of the fitted parameters. The
first observation is that (off-diagonal) correlations are generally
not very large. There is however one exception, \textit{i.e.} the
parameters $\sigma$ and $\lambda$ seem to be strongly
anti-correlated. This may indicate that the interplay between
$q$-Gaussian and Gaussian may be significantly $x$ dependent. We leave
a deeper study of this feature to a future publication.
\begin{table}[t]
  \centering
  \includegraphics[width=0.9\textwidth]{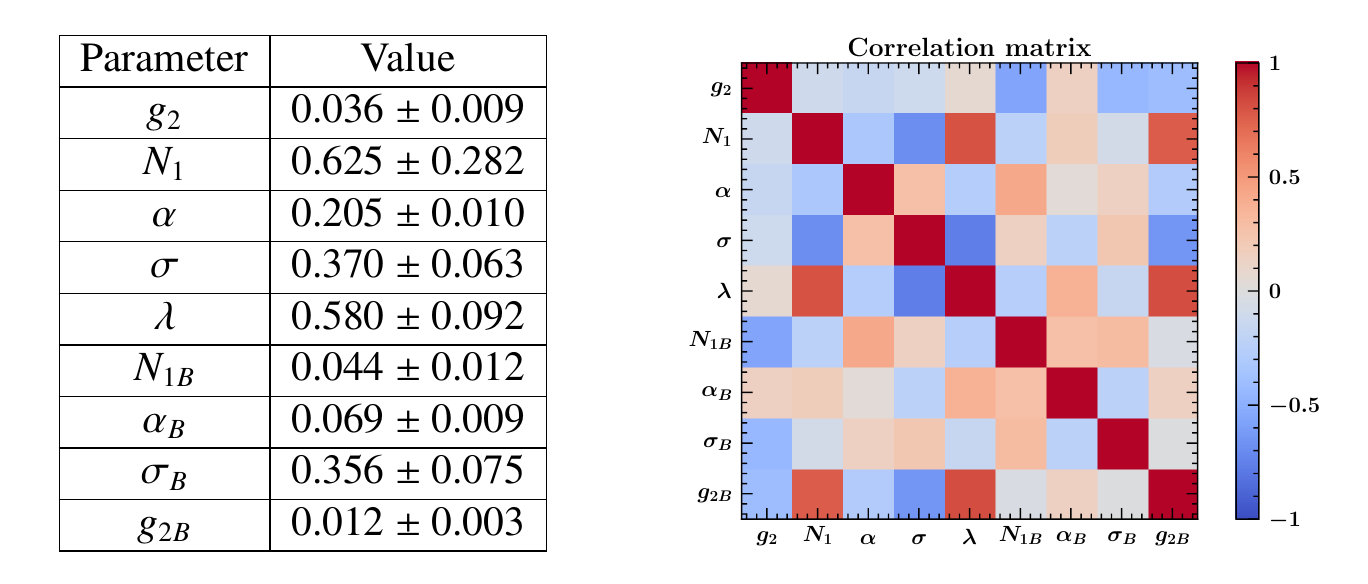}
\caption{Average and standard deviation over the Monte Carlo replicas
  of the free parameters fitted to the
  data and graphical representation of the correlation matrix.
  \label{t:fittedparameters}}
\end{table}


\begin{figure}[h!]
  \begin{centering}
    \includegraphics[width=0.49\textwidth]{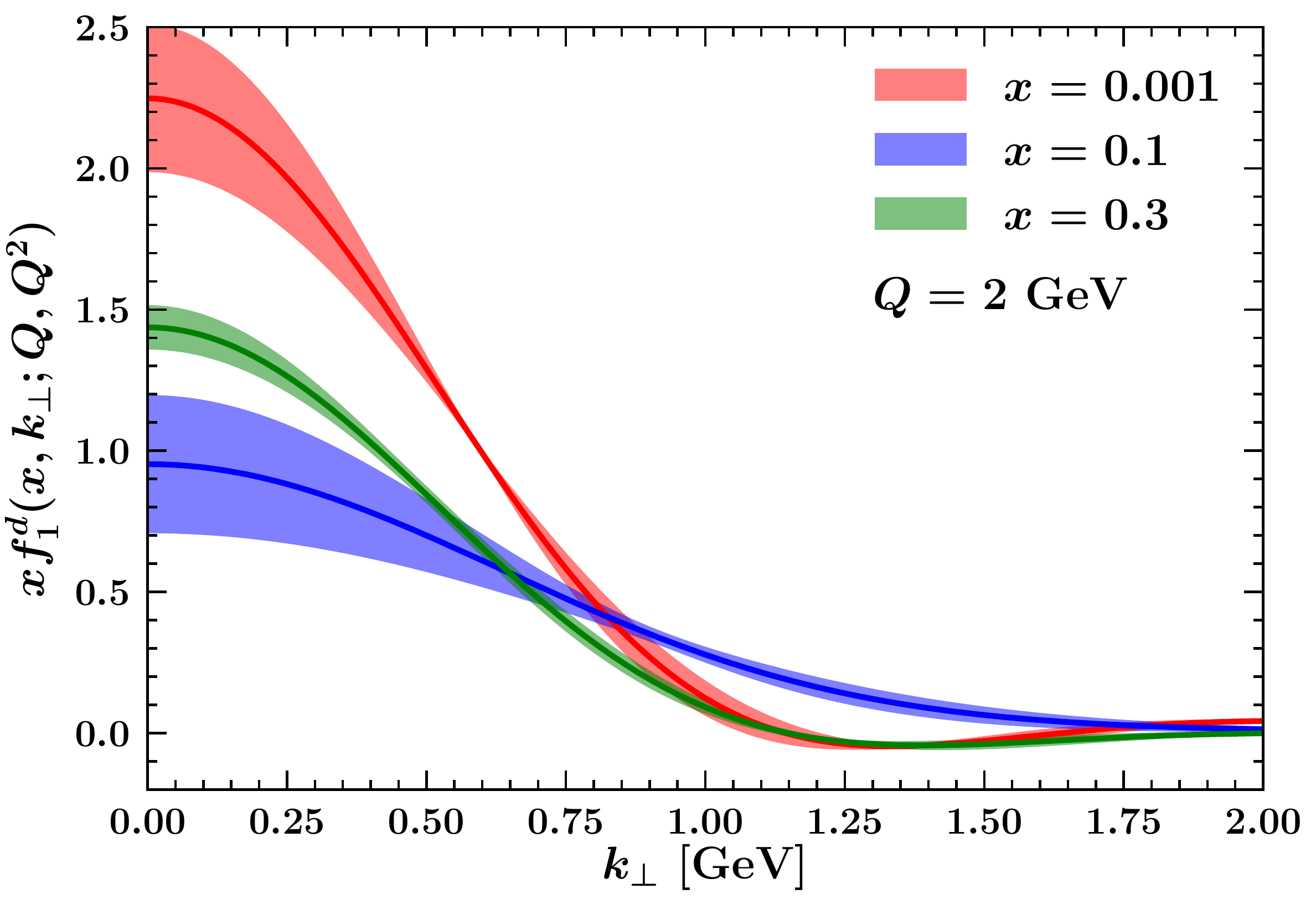}
    \includegraphics[width=0.49\textwidth]{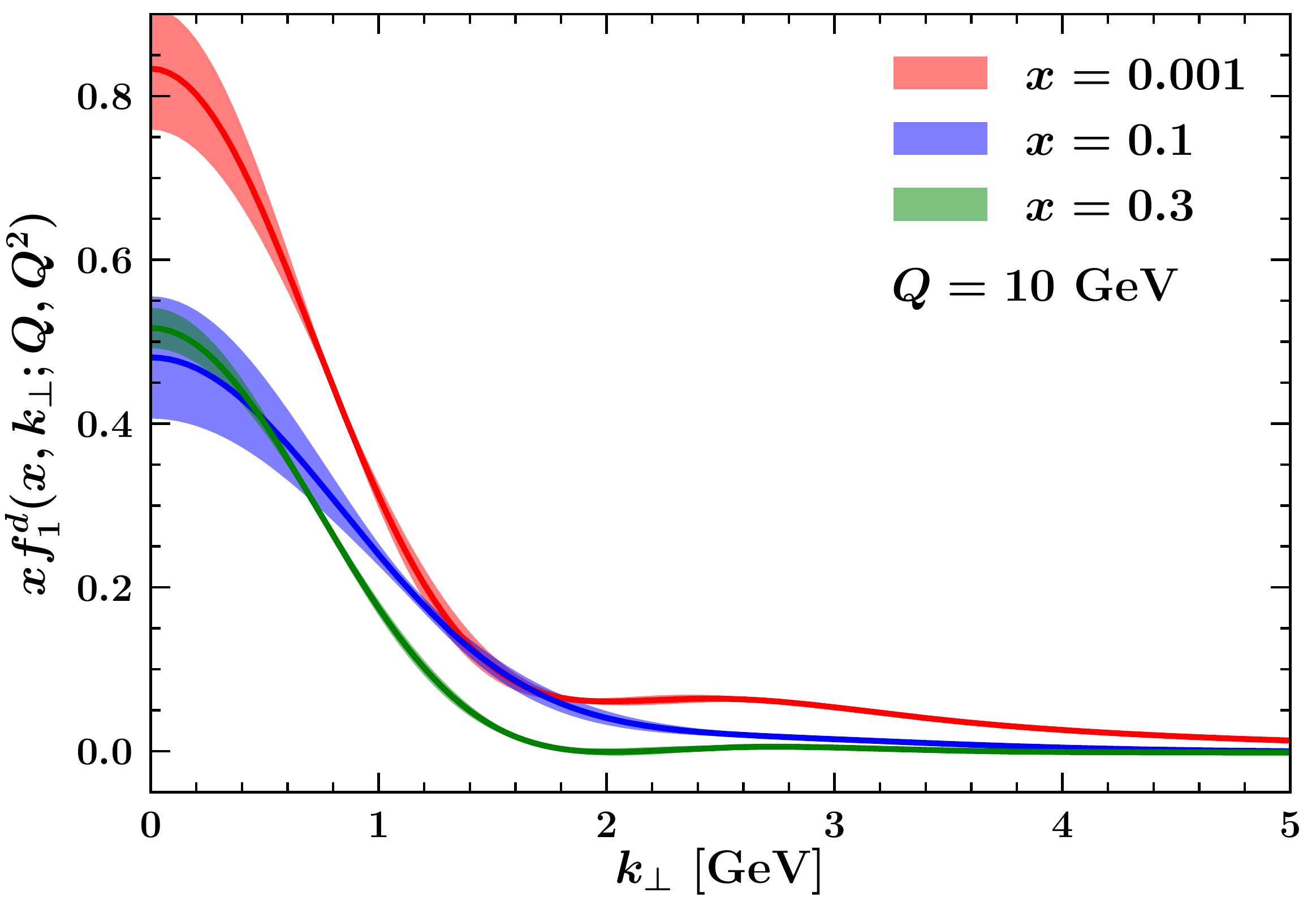}
    \caption{The TMD of the down quark at $\mu=\sqrt{\zeta}=Q=2$~GeV
      (left plot) and 10~GeV (right plot) as a function of the
      partonic transverse momentum $k_\perp$ for three different
      values of $x$. The bands give the 1-$\sigma$
      uncertainty.\label{fig:TMDs}}
  \end{centering}
\end{figure}
To conclude this section, in Fig.~\ref{fig:TMDs} we show the
down-quark TMD at $\mu=\sqrt{\zeta}=Q=2$~GeV (left plot) and 10~GeV
(right plot) as a function of the partonic transverse momentum
$k_\perp$ for $x=0.001, 0.1, 0.3$. The 1-$\sigma$ uncertainty bands
are also shown. As expected, TMDs are suppressed as $k_\perp$ grows
and the suppression becomes relatively stronger as $Q$ increases.

\subsection{Perturbative convergence}\label{ss:pertconv}

In the previous section we discussed the quality of our fit at
N$^3$LL, which is the best accuracy presently available. In this
section we show how the inclusion of perturbative corrections is
crucial to achieve a better description of the experimental data. To
this end, we performed fits at NLL$'$, NNLL, and NNLL$'$ (see
Sec.~\ref{s:pertordering}), and compared them to the N$^3$LL fit. We
did not consider LL and NLL accuracies because in both cases the
description of the data is very poor ($\chi^2\gtrsim 20$).

\begin{table}[h]
\centering
\begin{tabular}{|c|c|c|c|c|} \hline
& \quad NLL$'$ \quad & \quad NNLL \quad & \quad NNLL$'$ \quad & \quad N$^3$LL\quad \\
\hline
\vphantom{\bigg|}Global $\chi^2$ & 1126 & 571 & 379 & 360 \\
\hline
\end{tabular}
\caption{Values of the global $\chi^2$ of the fits at NLL$'$, NNLL, NNLL$'$, and N$^3$LL accuracy.\label{t:PerturbativeConvengence}}
\end{table}

\begin{figure}[h!]
  \begin{centering}
    \includegraphics[width=0.6\textwidth]{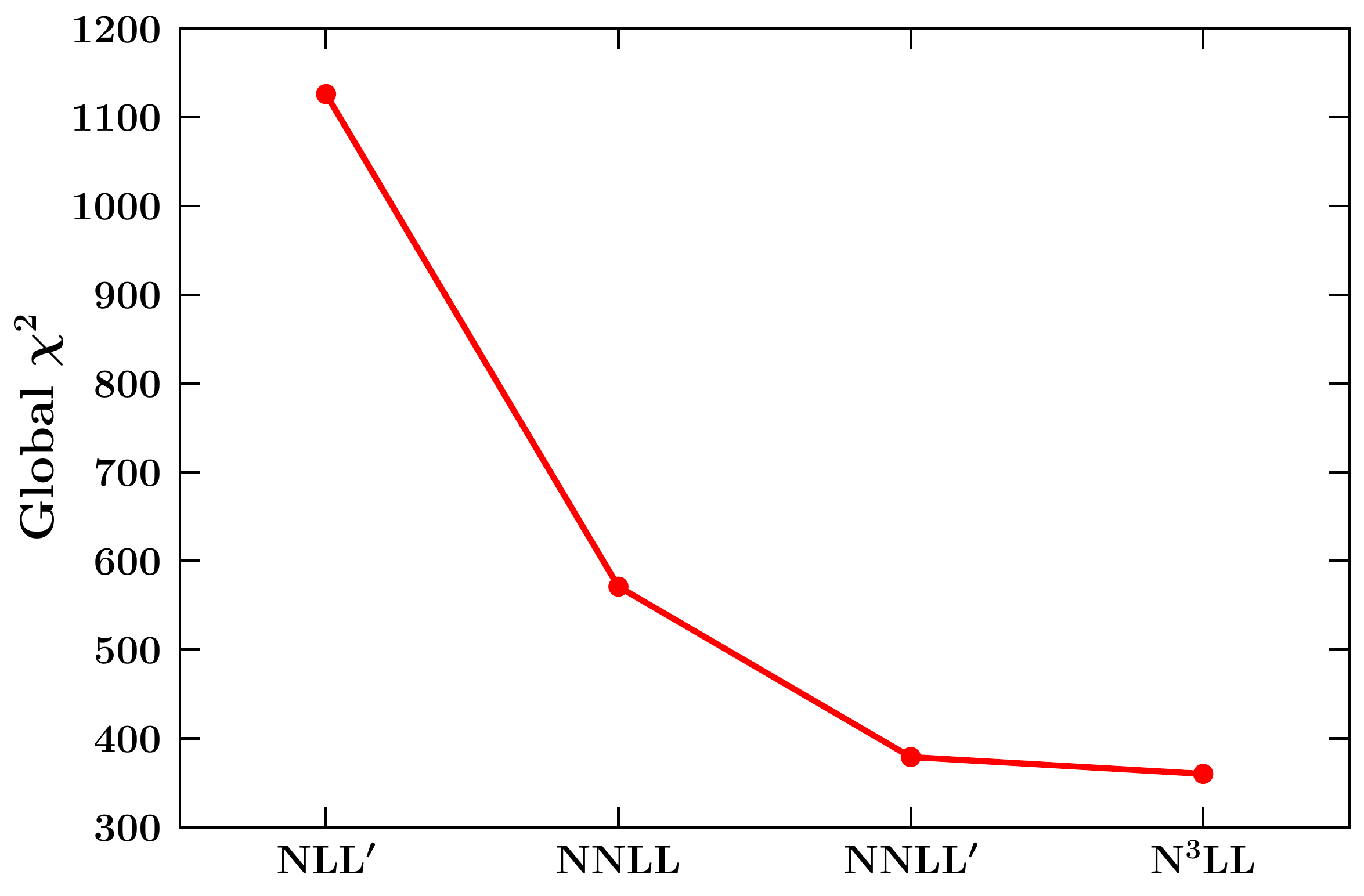}
    \caption{Graphical representation of
      Tab.~\ref{t:PerturbativeConvengence}.\label{fig:Convergence}}
  \end{centering}
\end{figure}
Tab.~\ref{t:PerturbativeConvengence} reports the values of the global
$\chi^2$ for each of the four accuracies considered. In order to
appreciate the significance of the differences,\footnote{Note that a
  difference of $n$ units at the level of the global $\chi^2$ roughly
  means a separation of around $\sqrt{n}$ standard deviations.} we
have reported the absolute values of the $\chi^2$ without dividing by
the number of data points $N_{\rm dat}$. Fig.~\ref{fig:Convergence}
shows a graphical representation of
Tab.~\ref{t:PerturbativeConvengence}.  The global quality of the fit
improves significantly as the perturbative accuracy increases. In
addition, Fig.~\ref{fig:Convergence} shows that the convergence rate
decreases when going to larger perturbative orders. On the one hand,
we conclude that it is necessary to include higher perturbative
corrections to obtain a good description of the data and that N$^3$LL
corrections are still significant. On the other hand, it appears that
the perturbative series is nicely converging and N$^3$LL accuracy
seems appropriate within the current experimental uncertainties.

\begin{figure}[t]
  \begin{centering}
    \includegraphics[width=0.75\textwidth]{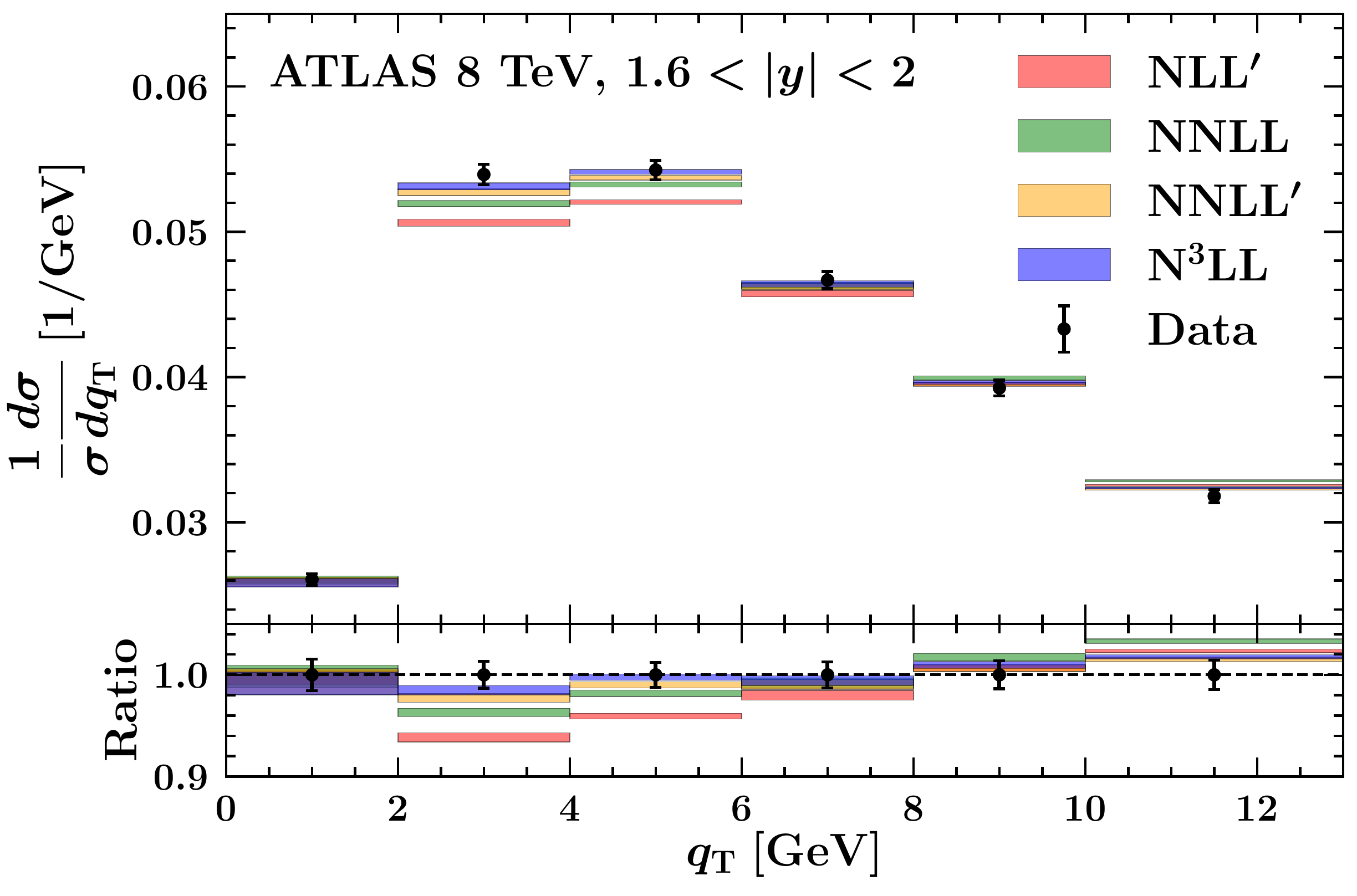}
    \caption{Comparison between experimental data for the ATLAS 8 TeV
      measurements in the bin 66 GeV $<Q<$ 116 GeV and $1.6 < |y| < 2$ and the
      theoretical predictions obtained from the fits to all
      perturbative orders considered in this analysis, \textit{i.e.}
      NLL$'$, NNLL, NNLL$'$, and N$^3$LL (see
      Sec.~\ref{s:pertordering}). The layout of the plot is the same
      as in Fig.~\ref{fig:DataTheoryComparison}.
      \label{fig:PerturbativeConvengence}}
  \end{centering}
\end{figure}
In order to quantify the numerical impact of higher-order corrections,
in Fig.~\ref{fig:PerturbativeConvengence} we compare the predictions
for all the available perturbative orders to the ATLAS 8~TeV data in
the bin 66 GeV $<Q<$ 116 GeV and $1.6 < |y| < 2$. This plot shows how
the inclusion of higher-order corrections improves the shape of the
predictions, particularly around the peak region.

\subsection{Reduced dataset and $x$ dependence}\label{ss:reddata}

The non-perturbative function $f_{\rm NP}$,
Eq.~(\ref{eq:separatation}), accounts for the large-$\bT$ behaviour of
TMDs. It is in general a function of $\bT$, $\zeta$, and $x$. While
the asymptotic dependence on $\bT$ is driven by first-principle
considerations (see Sec.~\ref{s:npfunc}) and the evolution with
$\zeta$ is determined by the Collins-Soper equation~(\ref{eq:eveqs}),
the dependence on $x$ is totally unknown. Moreover, a direct access to
the $x$ dependence is particularly difficult to achieve because it
requires cross-section data finely binned in rapidity $y$. In the
dataset considered here, only the ATLAS experiment delivers data
differential in rapidity. Therefore, one would expect that these
datasets provide most of the sensitivity to the $x$ dependence of
TMDs.

In order to test this conjecture, we employed a particularly simple
$x$-\textit{independent} parameterisation of the non-perturbative
function:
\begin{equation}\label{eq:DWS}
  f_{\rm NP}^{\rm DWS}(\bT,\zeta) =
  \exp\left[-\frac{1}{2}\left(g_1+g_2\ln\left(\frac{\zeta}{2 Q_0^2}\right)\right)\bT^2\right]\,,
\end{equation}
with two free parameters, $g_1$ and $g_2$, and $Q_0^2 = 1.6$~GeV$^2$
(inspired by the pioneering work of Davies, Webber, and
Stirling.~\cite{Davies:1984sp}). Using Eq.~(\ref{eq:DWS}) we first
performed a fit at N$^3$LL to the full dataset. Then we excluded the
ATLAS datasets differential in rapidity (but we kept the off-peak
ATLAS 8~TeV datasets because inclusive in rapidity).
\begin{table}[h]
\centering
\begin{tabular}{|c|c|c|} \hline
  &  Full dataset &  No $y$-differential data \\
  \hline
  \vphantom{\Big|}Global $\chi^2/N_{\rm dat}$ & 1.339 & 0.895 \\
  \hline
  \hline
  $g_1$ & 0.304 & 0.207 \\
  \hline
  $g_2$ & 0.028 & 0.093 \\
  \hline
\end{tabular}
\caption{The values of the global $\chi^2$ normalised to the number of
  data points $N_{\rm dat}$ from the fit to the full dataset and to a
  reduced dataset without the $y$-differential ATLAS datasets, both using the parameterisation in
  Eq.~(\ref{eq:DWS}). For completeness, we also report the best-fit values of the
  parameters $g_1$ and $g_2$.\label{t:xDependence}}
\end{table}
The resulting $\chi^2$s normalised to the number of data points are
reported in Tab.~\ref{t:xDependence}. For completeness, we also show
the best-fit values of the parameters $g_1$ and $g_2$.

Firstly, the $\chi^2$ of the fit to the full dataset using
Eq.~(\ref{eq:DWS}) (1.339) is significantly larger than that obtained
using the parameterisation in
Eqs.~(\ref{eq:fNPparam})-(\ref{eq:auxfuncs}) (1.020). This suggests
that an $x$-dependent $f_{\rm NP}$ is required to obtain a good
description of the data. Secondly, the $\chi^2$ of the fit without the
$y$-differential ATLAS data comes out to be particularly low
(0.895). We conclude that at N$^3$LL accuracy the $x$ dependence of
the TMDs extracted from the currently available DY data is mostly
constrained by the ATLAS data differential in the boson rapidity $y$.
We note however that the agreement with the very precise ATLAS data may be
influenced also by other small corrections (e.g. power corrections). 

\subsection{Dependence on the cut on $q_T/ Q$}\label{ss:cutscan}

As discussed in Sec.~\ref{s:theory}, our analysis is based on TMD
factorisation whose validity is restricted to the region $q_T\ll Q$.
As a consequence, we consider only measurements that respect this
constraint. More precisely, we require that the maximum value of the
ratio $q_T/Q$ for a point to be included in the fit be 0.2 (see
Sec.~\ref{s:data}). Despite this particular value seems to be
generally recognised in the literature (see, \textit{e.g.},
Ref.~\cite{Scimemi:2017etj}), it is interesting to study how the
global description of the dataset changes by varying this cut. This
will help us assess more quantitatively the validity range of TMD
factorisation.

\begin{figure}[t]
  \begin{centering}
    \includegraphics[width=0.6\textwidth]{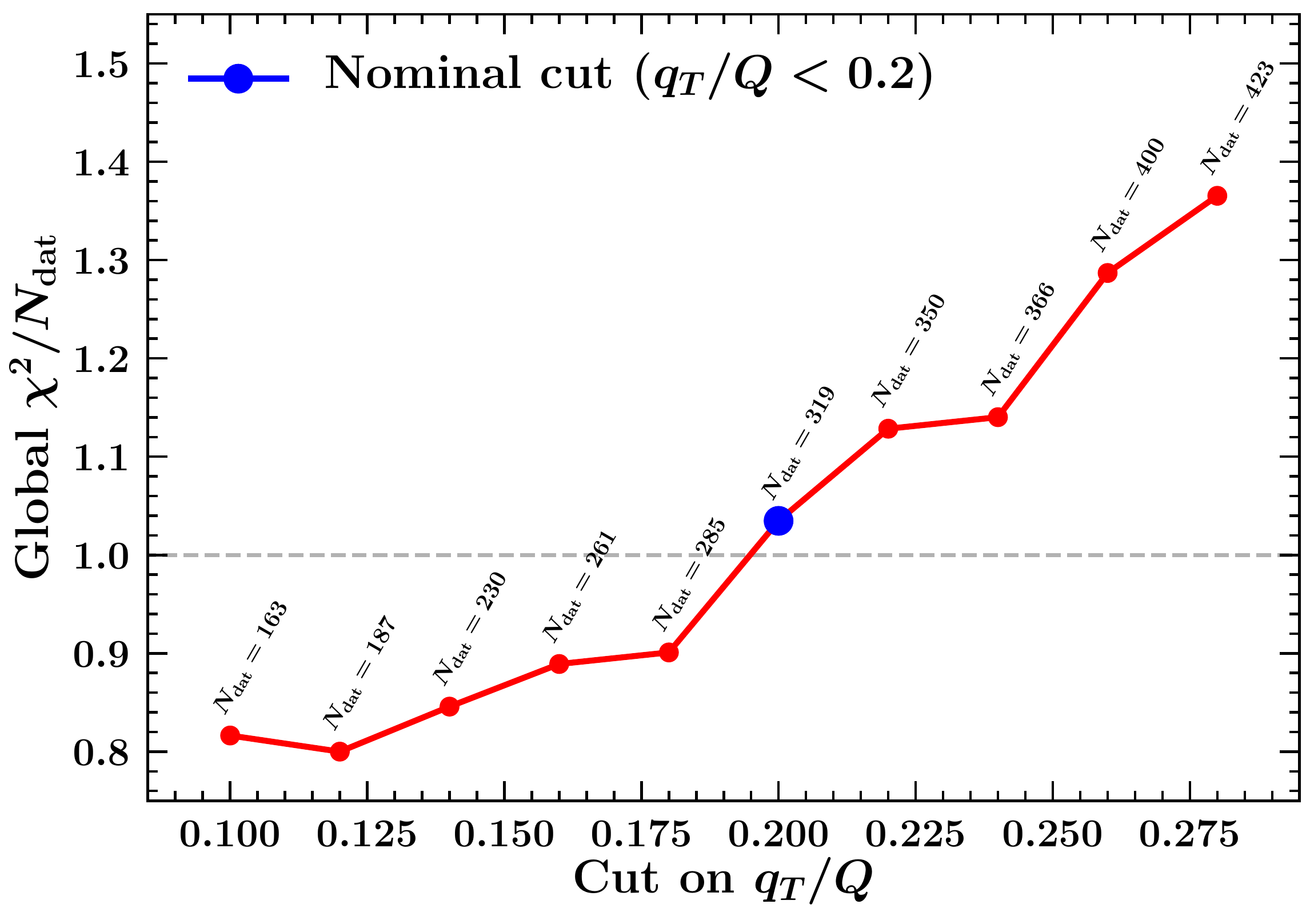}
    \caption{The global $\chi^2/N_{\rm dat}$ as a function of the cut
      on $q_T/Q$. The blue point corresponds to the reference cut used
      in this analysis.\label{fig:qToQScan}}
  \end{centering}
\end{figure}

Fig.~\ref{fig:qToQScan} displays the behaviour of the global
$\chi^2/N_{\rm data}$ for the N$^3$LL fit as a function of the $q_T/Q$
cut ranging between 0.1 and 0.28 in steps of 0.02. As expected, the
quality of the fit tends to degrade as the cut on $q_T/Q$
increases. Of course, it is impossible to draw a line between validity
and non-validity regions. However, this study gives a quantitative
justification for choosing the value 0.2 for the $q_T/Q$ cut.

\section{Conclusions}
\label{s:conc}

In this paper we presented an extraction of TMDs from Drell-Yan data
accurate up to N$^3$LL. The dataset used in this analysis includes
low-energy data from FNAL (E605 and E288) and RHIC (STAR) and
high-energy data from Tevatron (CDF and D0) and the LHC (LHCb, CMS,
and ATLAS), for a total of 353 data points.

The fit was performed with a proper treatment of the experimental
uncertainties, which were propagated into the fitted TMD distributions
by means of the Monte Carlo sampling method. This allowed us to obtain
a very good description of the entire dataset
($\chi^2/N_{\rm dat}=1.02$) without the need of introducing \textit{ad
  hoc} normalisations. A more detailed analysis of the fit quality
shows that both low- and high-energy datasets are separately well
described. This is a remarkable achievement given the very high
precision of the LHC datasets, especially those from ATLAS.

A particularly interesting aspect of our analysis concerns the QCD
convergence of the perturbative series. We performed fits at NLL$'$,
NNLL, NNLL$'$, and N$^3$LL accuracy and showed that the fit quality
improves significantly going from NLL$'$ to N$^3$LL. The difference
between the highest orders, \textit{i.e.} NNLL$'$ and N$^3$LL, is
moderate but still significant. This shows at the same time that the
perturbative series is converging, but also that N$^3$LL corrections
are relevant in relation to the current experimental uncertainties.

We parameterised the non-perturbative contributions by adopting a
reasonably flexible functional form: all nine free parameters turned out to
be well constrained, with moderate correlations amongst them.  An
important feature of our parameterisation of the non-perturbative
contribution $f_{\rm NP}$ is its explicit $x$ dependence. We proved
that the $x$-dependent part of $f_{\rm NP}$ is mostly constrained by
the rapidity-dependent on-peak data at 7 and 8~TeV from ATLAS. While
on the one hand, this was to be expected because the $x$ dependence is
strictly connected with the rapidity $y$, on the other hand it also
demonstrates that most of the datasets are not sensitive to the $x$
dependence of TMDs.

Finally, we studied the validity range of TMD factorisation in
Drell-Yan by varying the cut on $q_T/Q$. In line with the literature,
we found that the region $q_T \lesssim 0.2\,Q$ is appropriate when
working within the TMD factorisation framework.

In this paper we set the foundation for a number of future studies. In
the first place, we plan to extend the fitted dataset by including the
abundant and precise semi-inclusive DIS data from
HERMES~\cite{Airapetian:2012ki} and
COMPASS~\cite{Adolph:2013stb,Aghasyan:2017ctw}, as well as future data from
Jefferson Lab at 12 GeV~\cite{Dudek:2012vr}.
On top of providing access
to TMD fragmentation functions, we expect that the inclusion of semi-inclusive
DIS data will have an impact on the determination of the $x$ dependence of TMD
PDFs 
and
will make it possible
to determine the flavour dependence of the non-perturbative function
$f_{\rm NP}$.
We remark that a better knowledge of TMDs will be important not only to obtain a
deeper knowledge of hadron stucture and QCD, but also for precision studies in
high-energy processes involving hadrons, for instance for the
determination of critical Standard Model parameters such as the $W$
mass~\cite{Bacchetta:2018lna,Bozzi:2019vnl}.

In the future, the Electron-Ion Collider will provide an unprecedented
opportunity to make progress in the determination of
TMDs~\cite{Boer:2011fh,Accardi:2012qut}. Nevertheless, we are
convinced that the era of precision physics with TMDs has already
started and it will be beneficial also for studies at higher energies
in the perturbative domain of QCD.

\section*{Acknowledgments}

We thank P.~F.~Monni for discussions concerning the different
logarithmic orderings, and H. Avakyan for discussions on the 
functional forms for our parametrisation. This work is supported by
the European Research Council (ERC) under the European Union's Horizon
2020 research and innovation program (grant agreement No. 647981,
3DSPIN).

\appendix

\section{Numerics and delivery}\label{a:numerics}

In this appendix we give a brief general overview of the numerical
implementation of the analysis discussed above. The code used is
publicly available at
\begin{center}
\url{https://github.com/vbertone/NangaParbat}
\end{center}
where a more detailed documentation can be found along with a
collections of results. The code uses {\tt
  APFEL++}~\cite{Bertone:2013vaa,Bertone:2017gds} as an engine for the
computation of the theoretical predictions. In order to speed up the
fit on the non-perturbative function $f_{\rm NP}$,
Eq.~(\ref{eq:separatation}), we use interpolation techniques inspired
by those heavily used for collinear-factorisation
predictions~\cite{Kluge:2006xs,Britzger:2012bs,Carli:2010rw}. Schematically,
we reduce the computation of the cross section in
Eq.~(\ref{eq:phasespaceintegral}) for a given kinematic bin to the
weighted sum
\begin{equation}
  \frac{d\sigma}{dq_T} \simeq \sum_{n,\alpha,\tau} W_{n\alpha\tau} f_{\rm NP}(x_1^{(\alpha,\tau)},\bT^{(n)},\zeta^{(\tau)}) f_{\rm NP}(x_2^{(\alpha,\tau)},\bT^{(n)},\zeta^{(\tau)})\,,
\end{equation}
where the discrete variables $x_{1,2}^{(\alpha,\tau)}$, $\bT^{(n)}$,
and $\zeta^{(\tau)}$ run over appropriately defined grids. The
computationally expensive part of the calculation is isolated into the
weights $W_{n\alpha\tau}$ that are precomputed and stored. This
procedure makes the computation of predictions very fast and thus
suitable for a fit that requires a large number iterations.

In order to fit the function $f_{\rm NP}$ to data, we used two
independent codes: {\tt Minuit2}~\cite{James:1994vla} as implemented
in {\tt ROOT}, and {\tt ceres-solver}~\cite{ceres-solver}. While the
first ({\tt Minuit}) is routinely used for this kind of tasks since
many years, the second ({\tt ceres-solver}) is relatively new and
typically used for more complex problems such as image recognition, 3D
modeling, etc.. Recently, the {\tt xFitter}
Collaboration~\cite{Alekhin:2014irh} has used {\tt ceres-solver} for
fitting collinear PDFs~\cite{xfitter-web}, showing that this tool is
suitable also for this kind of tasks. Having two independent codes
within the same framework turned out to be particularly useful to
cross check our results.

All the datasets included in this analysis, except the preliminary
STAR data, have been taken from the public {\tt HEPData}
repository~\cite{Maguire:2017ypu} in {\tt YAML} format and slightly
adapted to fit our needs.

Finally, we mention that the TMDs sets determined in this analysis
will be made publicly available also through the {\tt TMDplotter}
interface~\cite{Hautmann:2014kza}.

\section{Integrating over $q_T$}\label{a:qTintegration}

Experimental measurements of differential distributions are usually
delivered as integrated over finite regions of the final-state
kinematic phase space (see Eq.~(\ref{eq:phasespaceintegral})). As a
consequence, in order to compare theoretical predictions to data, it
is necessary to carry out these integrations. These nested integrals,
if evaluated numerically, represent a heavy task that makes an
extraction of TMDs from Drell-Yan data computationally very intensive
and thus slow. While the integrals over $Q$ and $y$ do need to be
computed numerically, the integration in $q_T$ can be carried out
analytically which substantially reduces the numerical load.  To do
so, we exploit the following property of the Bessel functions
\begin{equation}
\frac{d}{dx}\left[x^n J_n(x)\right]=x^nJ_{n-1}(x)\,,
\end{equation}
that leads to
\begin{equation}\label{eq:besselproperty}
\int dx\,x J_0(x) = xJ_1(x)\quad\Rightarrow\quad \int_{x_1}^{x_2}
dx\,x J_0(x) = x_2J_1(x_2) - x_1J_1(x_1)\,.
\end{equation}
Neglecting for the moment the dependence on $q_T$ of the phase-space
reduction factor $\mathcal{P}$ (which is strictly correct for
inclusive observables in the final-state leptons), the differential
cross section in Eq.~(\ref{eq:crosssection}) has the following
structure
\begin{equation}
  \frac{d\sigma}{dQ dy dq_T} = \int_0^\infty d\bT\,S(\bT)\, q_T  J_0(\bT q_T)
\end{equation}
where $S$ is a function that depends on $\bT$ (and on the other
kinematic variables) but not on $q_T$. Using
Eq.~(\ref{eq:besselproperty}), one finds
\begin{equation}
\begin{array}{rcl}
\displaystyle \int_{q_{T,\rm min}}^{q_{T,\rm
  max}}dq_T\left[\frac{d\sigma}{dQ dy dq_T} \right] &=&\displaystyle \int_0^\infty d\bT\,S(\bT)\,
  \int_{q_{T,\rm min}}^{q_{T,\rm
  max}} dq_T\,   q_T J_0(\bT q_T) \\
\\
&=&\displaystyle\int_0^\infty d\bT\frac{S(\bT)}{\bT}\left[q_{T,\rm
  max}J_1(\bT q_{T,\rm max}) - q_{T,\rm
  min}J_1(\bT q_{T,\rm min})\right]\,.
\end{array}
\end{equation}
In conclusion, the quantity
\begin{equation}
K(q_T) \equiv \int_0^\infty d\bT\frac{S(\bT)}{\bT}\, q_T  J_1(\bT q_T)\,,
\end{equation}
is the indefinite integral over $q_T$ (the primitive function) of the
cross section in Eq.~(\ref{eq:crosssection}). Analogously to the
unintegrated cross section, $K$ can be computed numerically by
performing a Bessel transform of degree one rather than degree
zero. Therefore, the integral over a $q_T$ bin can be evaluated by
taking the difference of $K$ computed at the bin bounds:
\begin{equation}\label{eq:primitive}
  \int_{q_{T,\rm min}}^{q_{T,\rm
      max}}dq_T\left[\frac{d\sigma}{dQ dy dq_T} \right] = K(q_{T,\rm max})
  - K(q_{T,\rm min})\,,
\end{equation}
which is enormously more convenient than computing the integral
numerically.

\subsection{Kinematic cuts}\label{sec:kincuts}

In the presence of kinematic cuts, such as those on the final-state
leptons, the analytic integration over $q_T$ discussed above cannot be
directly performed. The reason is that the implementation of these
cuts effectively introduces the $q_T$-dependent function $\mathcal{P}$
in the integral
\begin{equation}
  \frac{d\sigma}{dQ dy dq_T} = \int_0^\infty d\bT\,S(\bT) \mathcal{P}(q_T) q_T  J_0(\bT q_T)\,,
\end{equation}
that prevents the direct use of
Eq.~(\ref{eq:besselproperty}). Fortunately, $\mathcal{P}$ is a
slowly-varying function of $q_T$ over the typical bin size. This
allows one to approximate the integral over the bins in $q_T$ as
\begin{equation}\label{eq:intbyparts}
\begin{array}{c}
\displaystyle  \int_{q_{T,\rm min}}^{q_{T,\rm
      max}} dq_T\, q_T J_0(\bT q_T) \mathcal{P}(q_T) \simeq \displaystyle
  \mathcal{P}\left(\frac{q_{T,\rm max}+q_{T,\rm min}}2\right)\int_{q_{T,\rm min}}^{q_{T,\rm
      max}} dq_T\, q_T J_0(\bT q_T) \\
\\
\displaystyle = \mathcal{P}\left(\frac{q_{T,\rm max}+q_{T,\rm
    min}}2\right) \frac{1}{\bT}\left[q_{T,\rm max} J_1(\bT q_{T,\rm max}) - q_{T,\rm min} J_1(\bT q_{T,\rm min}) \right]\,.
\end{array}
\end{equation}
Unfortunately, this structure is inconvenient because it mixes
different bin bounds and prevents a recursive computation.  However,
it is possible to go further and, assuming that the bin width is small
enough, we expand $\mathcal{P}$ in the following two equivalent ways
\begin{equation}\label{eq:expansions}
\displaystyle\mathcal{P}\left(\frac{q_{T,\rm max}+q_{T,\rm
    min}}2\right) = \left\{
\begin{array}{l}
\mathcal{P}\left(q_{T,\rm min}+\Delta q_T\right) \simeq
  \mathcal{P}\left(q_{T,\rm min}\right) + \mathcal{P}'\left(q_{T,\rm
  min}\right)\Delta q_T\\
\mathcal{P}\left(q_{T,\rm max}-\Delta q_T\right) \simeq
  \mathcal{P}\left(q_{T,\rm max}\right) - \mathcal{P}'\left(q_{T,\rm
  max}\right)\Delta q_T 
\end{array}\right.\,,
\end{equation}
with
\begin{equation}\label{eq:halfqTinterval}
\Delta q_T = \frac{q_{T,\rm max}- q_{T,\rm min}}2\,.
\end{equation}
Plugging the expansions above into Eq.~(\ref{eq:intbyparts}), one
finds
\begin{equation}\label{eq:lastexpP}
\begin{array}{rcl}
  \displaystyle  \bT\int_{q_{T,\rm min}}^{q_{T,\rm
  max}} dq_T\, q_T J_0(\bT q_T) \mathcal{P}(q_T)&\simeq&\displaystyle
                                                      q_{T,\rm
                                                      max}
                                                      J_1(\bT q_{T,\rm
                                                      max})\left[
                                                      \mathcal{P}\left(q_{T,\rm
                                                      max}\right)
                                                      - 
                                                      \mathcal{P}'\left(q_{T,\rm
                                                      max}\right)\Delta
                                                      q_T\right]\\
&-&\displaystyle q_{T,\rm
                                                      min}
                                                      J_1(\bT q_{T,\rm
                                                      min})\left[
                                                      \mathcal{P}\left(q_{T,\rm
                                                      min}\right)
                                                      +
                                                      \mathcal{P}'\left(q_{T,\rm
                                                      min}\right)\Delta
                                                      q_T\right]\,.
\end{array}
\end{equation}
The advantage of this formula as compared to Eq.~(\ref{eq:intbyparts})
is that each of the terms in the r.h.s. depends on one single bin
bound in $q_T$ rather than on a combination of two consecutive
bounds. This allows for a recursive computation of predictions in
neighbouring bins in $q_T$.

\section{Cuts on the final-state leptons}\label{a:LeptonCuts}

In this section, we derive the explicit expression of the phase-space
reduction factor $\mathcal{P}$ introduced in Sec.~\ref{s:theory}. This
factor is defined as\footnote{In Eq.~(\ref{eq:PSredDef}) a
  parity-violating term is neglected. We will argue in
  Sec.~\ref{as:pvcontr} that its contribution is negligible for
  realistic cuts.}
\begin{equation}\label{eq:PSredDef}
  \mathcal{P}(q) = \frac{\displaystyle \int_{\mbox{\footnotesize fid.
        reg.}}d^4p_1 d^4p_2 \,\delta(p_1^2) \delta(p_2^2)\theta(p_{1,0}) \theta(p_{2,0})\delta^{(4)}(p_1+p_2-q) L_\perp(p_1,p_2)}{\displaystyle \int d^4p_1 d^4p_2\, \delta(p_1^2) \delta(p_2^2) \theta(p_{1,0}) \theta(p_{2,0})\delta^{(4)}(p_1+p_2-q) L_\perp(p_1,p_2)}\,,
\end{equation}
where $p_1$ and $p_2$ are the four-momenta of the outgoing
leptons. The integral in the nu\-me\-ra\-tor extends over the fiducial
region defined by the cuts on the final-state leptons. The quantity
$L_\perp$ is defined as
\begin{equation}\label{eq:LT}
  L_\perp = g_\perp^{\mu\nu}L_{\mu\nu}\,,
\end{equation}
where $L_{\mu\nu}$ is the (parity-conserving part of the)
\textit{leptonic tensor} that, assuming massless leptons, reads
\begin{equation}\label{eq:lepttens}
L^{\mu\nu} = 4(p_1^{\mu}p_2^{\nu}+p_2^{\mu}p_1^{\nu}-g^{\mu\nu}p_1p_2)\,,
\end{equation}
while the transverse metric is given by
\begin{equation}\label{eq:transmetric}
  g_\perp^{\mu\nu} = g^{\mu\nu}+z^\mu z^\nu-t^\mu t^\nu\,.
\end{equation}
The vectors $z^\mu$ and $t^\mu$, in the Collins-Soper frame, are
defined as
\begin{equation}\label{eq:auxvects}
z^\mu = (\sinh y,\mathbf{0},\cosh y)\,,\qquad t^\mu = \frac{q^\mu}{Q}\,,
\end{equation}
and they are such that $z^2=-1$, $t^2=1$ and $(z\cdot q) = 0$. The
effect of integrating over the fiducial region in the numerator of
Eq.~(\ref{eq:PSredDef}) can be implemented by defining a generalised
$\theta$-function, $\Phi(p_1,p_2)$, that is equal to one inside the
fiducial region and zero outside. This allows one to integrate also
the numerator over the full phase-space of the two outgoing
leptons. Next, we integrate out one of the momenta, say $p_2$,
exploiting the momentum-conservation $\delta$-function:
\begin{equation}\label{eq:PSredDef2}
P(q)=\frac{\displaystyle \int d^4p \delta(p^2)
\delta((q-p)^2) \theta(p_{0})
  \theta(q_0-p_{0})L_\perp(p,q-p)\Phi(p,q-p)}{\displaystyle \int d^4p \delta(p^2)
\delta((q-p)^2) \theta(p_{0})
  \theta(q_0-p_{0})L_\perp(p,q-p)}\,,
\end{equation}
where we have renamed $p=p_1$. The remaining $\delta$-functions can be
used to constrain two of the four components of the momentum $p$. The
first, $\delta(p^2)$, is typically used to set the energy component of
$p$, $p_0$, on the mass shell. Since the leptons are massless, this
produces
\begin{equation}\label{eq:phasespacemeasure}
\int d^4p\delta(p^2)\theta(p_0) = \int d^4p\,\delta(p_0^2-|\mathbf{p}|^2)\theta(p_0)=\int\frac{dp_0d^3\mathbf{p}}{2|\mathbf{p}|}\delta(p_0-|\mathbf{p}|)=\int\frac{d^3\mathbf{p}}{2|\mathbf{p}|}\,.
\end{equation}
Of course, the four-momentum $p$ appearing in the rest of the
integrand has to be set on shell ($p_0=|\mathbf{p}|$). Now we express
the three-dimensional measure $d^3\mathbf{p}$ in terms of the
transverse momentum $\mathbf{p}_T$, the pseudo-rapidity $\eta$, and
the azimuthal angle $\phi$ of the lepton:
\begin{equation}
\int\frac{d^3\mathbf{p}}{2 |\mathbf{p}|} =  \int \frac{d|\mathbf{p}_T|^2}{4}
d\eta\,d\phi\,.
\end{equation}
Now we consider the second $\delta$-function, $\delta((q-p)^2)$, in
Eq.~(\ref{eq:PSredDef2}). It is convenient to express the vectors $q$
and $p$ in terms of the respective invariant mass, pseudo-rapidity,
and transverse momentum:
\begin{equation}\label{eq:qpexplicit}
\begin{array}{rcl}
q&=&\left(M\cosh y,\mathbf{q}_T,M\sinh y\right)\,,\\
p&=&\left(|\mathbf{p}_T|\cosh\eta,\mathbf{p}_T,|\mathbf{p}_T|\sinh\eta\right)\,,
 \end{array}
\end{equation}
with $M=\sqrt{Q^2+|\mathbf{q}_T|^2}$. Without loss of generality, we
assume that the two-dimensional vector $\mathbf{q}_T$ is aligned with
the $x$ axis so that
$\mathbf{p}_T\cdot \mathbf{q}_T =
|\mathbf{p}_T||\mathbf{q}_T|\cos\phi$.\footnote{In
  the general case in which $\mathbf{q}_T$ forms an angle $\beta$ with
  the $x$ axis, the scalar product would result in
  $|\mathbf{p}_T||\mathbf{q}_T|\cos(\phi-\beta)$. However, for
  observables inclusive in azimuthal angle, the angle $\beta$ can
  always be reabsorbed in a redefinition of $\phi$.} This leads to
\begin{equation}
\delta((q-p)^2) = \delta\left(Q^2-2 |\mathbf{p}_T|\left[M\cosh\left(\eta - y\right)-|\mathbf{q}_T|\cos\phi\right]\right)\,,
\end{equation}
so that
\begin{equation}\label{eq:LT2}
\mathcal{P}(q)=\frac{\displaystyle
  \int \frac{d|\mathbf{p}_T|^2}{4} d\eta\,d\phi\,\delta\left(Q^2-2 |\mathbf{p}_T|\left[M\cosh\left(\eta - y\right)-|\mathbf{q}_T|\cos\phi\right]\right)L_\perp(p,q-p)\Phi(p,q-p)}{\displaystyle
 \int \frac{d|\mathbf{p}_T|^2}{4} d\eta\,d\phi\,\delta\left(Q^2-2 |\mathbf{p}_T|\left[M\cosh\left(\eta - y\right)-|\mathbf{q}_T|\cos\phi\right]\right)L_\perp(p,q-p)}\,,
\end{equation}
where the vector $p$ is understood to be on-shell. Now we compute
$L_\perp(p,q-p)$ contracting $L_{\mu\nu}$ in Eq.~(\ref{eq:lepttens})
with the transverse metric $ g_\perp^{\mu\nu}$ in
Eq.~(\ref{eq:transmetric}) using Eq.~(\ref{eq:qpexplicit}):
\begin{equation}
  L_\perp(p,q-p) = 2Q^2\left[1+4 \sinh^2(y-\eta)\frac{|\mathbf{p}_T|^2}{Q^2}\right]\,.
\end{equation}
We can now integrate out one of the variables in the integrals in
Eq.~(\ref{eq:LT2}) by making use of the remaining $\delta$-function.
Somewhat counterintuitively, it is convenient to integrate over
$|\mathbf{p}_T|$. This produces
\begin{equation}\label{eq:LT3}
  P(q)=\frac{\displaystyle \int_{-\infty}^{\infty} d\eta\int_{0}^{2\pi}d\phi\,\left[\frac{2\overline{p}_T^2}{Q^2}+
  2\sinh^2(y-\eta)\frac{\overline{p}_T^4} {Q^4}\right]\Phi(\overline{p},q-\overline{p})}{\displaystyle \int_{-\infty}^{\infty} d\eta\int_{0}^{2\pi}d\phi\,\left[\frac{2\overline{p}_T^2}{Q^2}+
  2\sinh^2(y-\eta)\frac{\overline{p}_T^4} {Q^4}\right]}\,,
\end{equation}
where $\overline{p}_T$ is defined as
\begin{equation}\label{eq:overpT}
  \overline{p}_T =\frac{Q^2}{2 |\mathbf{q}_T|}\frac1{\left[\frac{M\cosh\left(\eta - y\right)}{|\mathbf{q}_T|}-\cos\phi\right]}\,.
\end{equation}
and $\overline{p}$ symbolises the on-shell vector $p$ with the
absolute value of the transverse component set equal to
Eq.~(\ref{eq:overpT}). Next we turn to consider the integral in
$\phi$. To this end, the following relation
\begin{equation}\label{eq:intoverphi}
\int_0^{2\pi}d\phi\, f(\cos\phi) = \int_{-1}^1\frac{dx}{\sqrt{1-x^2}}\left[f(x)+f(-x)\right]\,,
\end{equation}
along with the \textit{indefinite} integrals
\begin{equation}\label{eq:complicatedintegral}
\int \frac{dx}{(a\pm
  x)^2\sqrt{1-x^2}}=\frac{\sqrt{1-x^2}}{(a^2-1)(x\pm
  a)}\pm\frac{a}{(a^2-1)^{3/2}}\tan^{-1}\left(\frac{1\pm ax}{\sqrt{a^2-1}\sqrt{1-x^2}}\right)\,,
\end{equation}
and
\begin{equation}\label{eq:complicatedintegral2}
\begin{array}{rcl}
\displaystyle\int \frac{dx}{(a\pm
  x)^4\sqrt{1-x^2}}&=&\displaystyle\frac{\sqrt{1-x^2}\left[(11a^2+4)x^2\pm
                       3 a(9a^2+1)x + (18a^4-5a^2+2)\right]}{6(a^2-1)^3(x\pm
  a)^3}\\
\\
&\pm&\displaystyle\frac{a(2a^2+3)}{2(a^2-1)^{7/2}}\tan^{-1}\left(\frac{1\pm
      ax}{\sqrt{a^2-1}\sqrt{1-x^2}}\right)\,,
\end{array}
\end{equation}
enable us to compute analytically the primitive function of the
integrals in $\phi$ in
Eq.~(\ref{eq:LT3}). Eqs.~(\ref{eq:complicatedintegral})
and~(\ref{eq:complicatedintegral2}) are particularly useful because
they allow us to compute the integral over $\phi$ analytically also in
the presence of cuts. Let us first compute the integral in the
denominator of Eq.~(\ref{eq:LT3}), \textit{i.e.} the integral of
$L_\perp$ over the full phase-space. To do so, using
Eqs.~(\ref{eq:complicatedintegral})
and~(\ref{eq:complicatedintegral2}), we compute the following definite
integrals
\begin{equation}\label{eq:defintoverphi}
\int_{-1}^{1} \frac{dx}{(a\pm x)^2\sqrt{1-x^2}}=\frac{\pi a}{(a^2-1)^{3/2}}\,,
\end{equation}
and:
\begin{equation}\label{eq:defintoverphi2}
\int_{-1}^{1} \frac{dx}{(a\pm
  x)^4\sqrt{1-x^2}}=\frac{\pi a(2a^2+3)}{2(a^2-1)^{7/2}}\,.
\end{equation}
Using these results, and finally integrating over $\eta$, gives the
well-known result
\begin{equation}
\int d^4p_1 d^4p_2\, \delta(p_1^2) \delta(p_2^2) \theta(p_{1,0})
\theta(p_{2,0})\delta^{(4)}(p_1+p_2-q) L_\perp(p_1,p_2) = \frac{4\pi}{3}Q^2\,.
\end{equation}

In order to compute the numerator of Eq.~(\ref{eq:LT3}), we need to
insert the appropriate function $\Phi$. Typically, in DY production
the kinematic cuts are imposed independently on the same variables for
both the final-state leptons.  Therefore, the function $\Phi$
factorises into two identical functions acting on each lepton
momentum:
\begin{equation}
\Phi(p_1,p_2) = \Theta(p_1)\Theta(p_2)\,.
\end{equation}
We are specifically interested in kinematic cuts on the rapidity and
on the transverse momentum of the following kind
\begin{equation}
  \eta_{\rm
    min} < \eta_{1(2)} < \eta_{\rm max}\quad\mbox{and}\quad |\mathbf{p}_{T,1(2)}| > p_{T,\rm min}\,.
\end{equation}
Therefore
\begin{equation}
  \Theta(p) = \vartheta(\eta - \eta_{\rm min})\vartheta(\eta_{\rm max}-\eta) \vartheta(|\mathbf{p}_{T}| - p_{T,\rm min}) \,. 
\end{equation}
Using Eqs.~(\ref{eq:qpexplicit}) and~(\ref{eq:overpT}) gives
\begin{equation}\label{eq:almostfinal}
\begin{array}{rcl}
\Phi(p,q-p) &=&\displaystyle \vartheta(\eta-\eta_{\rm min}) \times
\vartheta(\eta_{\rm max}-\eta) \\
&\times& \vartheta(\cos\phi - f^{(2)}(\eta,
                p_{T,\rm min})) \\
&\times&\displaystyle
         \vartheta(f^{(3)}(\eta,\eta_{\rm min})-\cos\phi) \times \vartheta(f^{(3)}(\eta,\eta_{\rm max})-\cos\phi)\\
&\times&\vartheta(f^{(4)}(\eta,
         p_{T,\rm min})-\cos\phi)\,,
\end{array}
\end{equation}
with
\begin{equation}\label{eq:relevantfuncs}
\begin{array}{rcl}
f^{(2)}(\eta, p_{T,\rm cut}) & = &\displaystyle \frac{2M p_{T,\rm cut}\cosh(\eta-y) 
                    - Q^{2}}{2p_{T,\rm cut}
                    |\mathbf{q}_T|}\,, \\
\\
f^{(3)}(\eta,\eta_{\rm cut}) & = &\displaystyle \frac{M \cosh(\eta-y)}{|\mathbf{q}_T|
                    }-\frac{Q^{2} \left(\sinh(\eta
                                   -y)\coth(y-\eta_{\rm cut})+\cosh(\eta-y)\right)}{2|\mathbf{q}_T|  M}\,,\\
\\
f^{(4)}(\eta, p_{T,\rm cut}) & = &\displaystyle \frac{M \cosh(\eta-y)(Q^2 - 2
                    p_{T,\rm cut}^{2} + 2 |\mathbf{q}_T|^2)- Q^{2} \sqrt{M^{2} \sinh ^{2} (\eta-y) + p_{T,\rm cut}^{2} }}{2 |\mathbf{q}_T| \left(M^{2} - p_{T,\rm cut}^{2}\right)}\,.
\end{array}
\end{equation}
Now the question is identifying the integration domain on the
$(\eta,\cos\phi)$-plane defined by $\Phi(p,q-p)$ in
Eq.~(\ref{eq:almostfinal}). Considering that $-1\leq\cos\phi\leq 1$,
Eq.~(\ref{eq:almostfinal}) can be written in an more convenient way
as
\begin{equation}\label{eq:final}
\begin{array}{rcl}
\Phi(p,q-p) &=& \vartheta(\eta-\eta_{\rm min})\vartheta(\eta_{\rm
  max}-\eta)  \\
&\times&\vartheta(\cos\phi -
  \mbox{max}[f^{(2)}(\eta,p_{T,\rm min}),-1])\\
&\times&\vartheta(\mbox{min}[f^{(3)}(\eta,\eta_{\rm min}),f^{(3)}(\eta,\eta_{\rm
         max}), f^{(4)}(\eta,p_{T,\rm
         min}),1]-\cos\phi)\,.
\end{array}
\end{equation}
Now we use Eq.~(\ref{eq:intoverphi}) to change $\cos\phi$ into $x$.
This way, the double integral at the numerator of Eq.~(\ref{eq:LT3})
reads
\begin{equation}
\int_{-\infty}^{\infty}d\eta\int_{-1}^{1}dx\,\Phi(p,q-p)\dots =
\int_{\eta_{\rm min}}^{\eta_{\rm
    max}}d\eta\,\vartheta(x_2(\eta)-x_1(\eta))\int_{x_1(\eta)}^{x_2(\eta)}dx\dots\,.
\end{equation}
with
\begin{equation}
\begin{array}{rcl}
x_1(\eta) &=& \mbox{max}[f^{(2)}(\eta,p_{T,\rm min}),-1]\\
x_2(\eta) &=& \mbox{min}[f^{(3)}(\eta,\eta_{\rm min}),f^{(3)}(\eta,\eta_{\rm
         max}),f^{(4)}(\eta,p_{T,\rm
         min}),1]\,.
\end{array}
\end{equation}

\begin{figure}[t]
  \begin{centering}
    \includegraphics[width=0.8\textwidth]{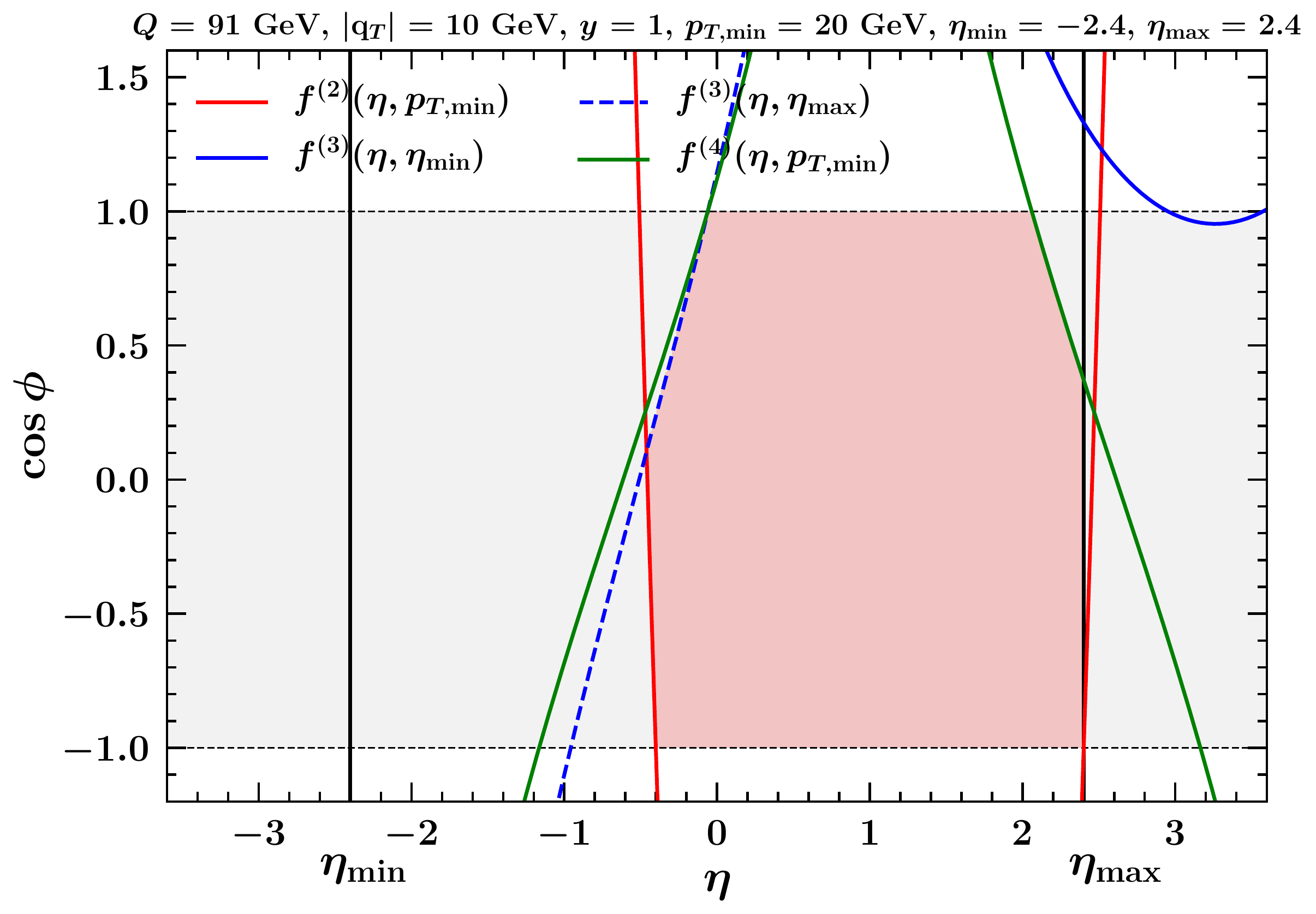}
    \caption{The red area indicates the integration domain of the
      numerator of the phase-space reduction factor Eq.~(\ref{eq:LT3})
      for $p_{T,\rm min}=20$~GeV and
      $-\eta_{\rm min}=\eta_{\rm max}=2.4$ at $Q=91$~GeV,
      $|\mathbf{q}_T|=10$ GeV, and $y=1$.\label{fig:IntDomain}}
  \end{centering}
\end{figure}
As an example, Fig.~\ref{fig:IntDomain} shows the integration domain
of the numerator of Eq.~(\ref{eq:LT3}) for $p_{T,\rm min}=20$~GeV and
$-\eta_{\rm min}=\eta_{\rm max}=2.4$ at $Q=91$~GeV,
$|\mathbf{q}_T|=10$~GeV, and $y=1$. The grey band corresponds to the
region $-1\leq\cos\phi\leq 1$. The $\theta$-functions in the first
line of Eq.~(\ref{eq:final}) limits the region to the vertical strip
defined by $\eta_{\rm min} < \eta < \eta_{\rm max}$ (black vertical
lines), the $\theta$-function in the second line defines the region
above the red line, finally the $\theta$-functions in the third line
defines the region below the blue and green lines. The intersection of
all regions gives the red-shaded area corresponding to the integration
domain.

Gathering all pieces, the final expression for the phase-space
reduction factor reads
\begin{equation}\label{eq:finalformula}
  \mathcal{P}(q)=\mathcal{P}(Q,y,q_T)=\displaystyle \int_{\eta_{\rm
      min}}^{\eta_{\rm
      max}}d\eta\,\vartheta(x_2(\eta)-x_1(\eta))\left[\overline{F}(x_2(\eta),\eta)-\overline{F}(x_1(\eta) ,\eta)\right]\,.
\end{equation}
The function $\overline{F}$ is given by the combination
\begin{equation}\label{eq:FGcombination}
\overline{F}(x,\eta) = \frac34 F(x,\eta)+ \frac14 G(x,\eta)\,,
\end{equation}
with
\begin{equation}
\begin{split} 
\label{eq:integrandF}
\displaystyle F(x ,\eta)&= \displaystyle \frac{1}{4\pi}\frac{Q^2}{E_q^2-q_T^2}\Bigg\{\frac{q_T^2 x\sqrt{1-x^2}}{x^2 q_T^2-
  E_q^2}\\
&- \displaystyle \frac{E_q}{\sqrt{E_q^2-q_T^2}}\left[\tan^{-1}\left(\frac{q_T-
    xE_q}{\sqrt{E_q^2-q_T^2}\sqrt{1-x^2}}\right)-\displaystyle\tan^{-1}\left(\frac{q_T+
    xE_q}{\sqrt{E_q^2-q_T^2}\sqrt{1-x^2}}\right)\right]\Bigg\}\,,
\end{split} 
\end{equation} 
and
\begin{equation}
\begin{split}
\label{eq:integrandG}
\displaystyle G(x ,\eta)&=\displaystyle 
                            \frac{1}{16\pi }\sinh^2(y-\eta)\frac{Q^4}{(E_q^2-q_T^2)^3}
                            \Bigg\{\sqrt{1-x^2}q_T\\
&\quad \times \displaystyle\Bigg[\frac{(11E_q^2q_T^2+4q_T^4)x^2+
                       3 E_qq_T(9E_q^2+q_T^2)x + (18E_q^4-5E_q^2q_T^2+2q_T^4)}{(xq_T+
  E_q)^3}\\
&\quad + \displaystyle\frac{(11E_q^2q_T^2+4q_T^4)x^2-
                       3 E_qq_T(9E_q^2+q_T^2)x + (18E_q^4-5E_q^2q_T^2+2q_T^4)}{(xq_T-
  E_q)^3}\Bigg]\\
&\quad - \displaystyle\frac{6E_q (2E_q^2+3q_T^2)}{\sqrt{E_q^2-q_T^2}}
        \Biggl[\tan^{-1}\left(\frac{q_T-
       xE_q}{\sqrt{E_q^2-q_T^2}\sqrt{1-x^2}}\right)
      \\
& \hspace{5cm}    -\tan^{-1}\left(\frac{q_T+
      xE_q}{\sqrt{E_q^2-q_T^2}\sqrt{1-x^2}}\right)\Biggr]
\Bigg\}\,,
\end{split} 
\end{equation} 
where we have defined $E_q = M\cosh(\eta-y)$ and $q_T=|\mathbf{q}_T|$.
Interestingly, in the limit $y=q_T=0$ and assuming
$\eta_{\rm min} = -\eta_{\rm max}$, $\mathcal{P}$ can be computed
analytically. The result is
\begin{equation}
  \mathcal{P}(Q,0,0)=\vartheta(Q- 2p_{T,\rm
    min})\tanh(\mbox{max}[\eta_{\rm max},\overline{\eta}])\left[1-\frac{1}{4\cosh^2(\mbox{max}[\eta_{\rm max},\overline{\eta}])}\right]\,,
\end{equation}
with $\overline{\eta}$ defined as
\begin{equation}\label{eq:etabardef}
\overline{\eta} =\cosh^{-1}\left(\frac{Q}{2p_{T,\rm min}}\right)\,.
\end{equation}
The relation above can be written more explicitly as
\begin{equation}\label{eq:partcase2}
{
\mathcal{P}(Q,0,0) = 
\left\{
\begin{array}{ll}
0 & \quad Q< 2p_{T,\rm min}\,,\\
 \left(1-\frac{p_{T,\rm min}^2}{Q^2}\right)\sqrt{1-\frac{4 p_{T,\rm min}^2}{Q^2}}& \quad 2p_{T,\rm min} \leq Q < 2p_{T,\rm min}\cosh\eta_{\rm max}\,,\\
 \tanh(\eta_{\rm max})\left[1-\frac{1}{4\cosh^2(\eta_{\rm max})}\right] & \quad Q \geq 2p_{T,\rm min}\cosh\eta_{\rm max}\,.
\end{array}
\right.}
\end{equation}

\subsection{Azimuthally-dependent contributions}\label{as:pvcontr}

Azimuthally-dependent modulations disappear in the cross sections
if the integration over the azimuthal angle of the virtual boson, $\Phi$, is
complete. 
In the presence of cuts on the final-state leptons, these modulations could
generate contributions that were neglected in our analysis,
but could be relevant for the description of high-precision data.

We first consider 
parity-violating effects that generate a $\sin \Phi$
modulation~\cite{Boer:1999mm}. 
These contributions stem from
interference of the \textit{antisymmetric} contributions to the lepton
tensor, proportional to
$p_1^{\mu} p_2^{\nu}\epsilon_{\mu\nu\rho\sigma}$, and to the hadronic
tensor, proportional to $\epsilon_{\perp}^{\mu\nu}$ defined as
\begin{equation}
\epsilon_{\perp}^{\mu\nu}\equiv \epsilon^{\mu\nu\rho\sigma}t_\rho z_\sigma\,,
\end{equation}
where $t^\mu$ and $z^\mu$ are given in Eq.~(\ref{eq:auxvects}).
Therefore, the contributions we are after result from the contraction
of the following Lorentz structures
\begin{equation}\label{eq:pvphasespace}
L_{\rm PV}\equiv p_1^{\mu}
p_2^{\nu}\epsilon_{\mu\nu\rho\sigma}\epsilon_{\perp}^{\rho\sigma}=\frac{2|\mathbf{p}_T|^2}{Q}\sinh(y-\eta)\left[M\cosh(y-\eta)-|\mathbf{q}_T|\cos\phi\right]\,.
\end{equation}
Due to the presence of $\sinh(y-\eta)$, Eq.~(\ref{eq:pvphasespace}) is
such that
\begin{equation}\label{eq:noPVcontr}
\int_{-\infty}^{\infty} d\eta\,L_{\rm PV} = 0\,.
\end{equation}
Therefore, for observables inclusive in the lepton phase space, the
parity-violating term does not give any contribution. Conversely, the
presence of cuts on the final-state leptons may prevent
Eq.~(\ref{eq:noPVcontr}) from being satisfied, leaving a residual
contribution. In order to quantify this effect, we have taken the same
steps performed above to integrate $L_{\rm PV}$ over the fiducial
region. It turns out that, for realistic cuts, the numerical size of
$\mathcal{P}_{\rm PV}$ relative to the parity-conserving $\mathcal{P}$
is never larger than $\mathcal{O}(10^{-6})$. We conclude that the impact of
parity-violating effects in the present analysis is negligible.

Finally, we consider also $\cos \Phi$ modulations,  stemming from
the following contraction:
\begin{equation}
  L_\phi=(z^\mu t^\nu+z^\nu t^\mu)L_{\mu\nu}\,,
\end{equation}
where the (symmetric part of the) leptonic tensor reads:
\begin{equation}
L^{\mu\nu} =
4(p_1^{\mu}p_2^{\nu}+p_2^{\mu}p_1^{\nu}-g^{\mu\nu}p_1p_2)\,.
\end{equation}
We find that
\begin{equation}
L_\phi=16 \frac{p_T^2}{Q}\sinh(y-\eta)\left[\frac{Q^2}{2 p_T
  }-M\cosh(y-\eta)+q_T\cos\phi\right]\,.
\end{equation}
Due to the presence of the overall factor $\sinh(y-\eta)$, for
relatively central rapidities and for symmetric cuts this term is
expected to be very small, in particular to be comparable in size to
the parity violating contribution. Moreover, this term would be multiplied by
a structure function that has been measured to be small, below 4\% in the
region of interest here~\cite{Aad:2016izn}.


\bibliographystyle{JHEP}
\bibliography{PV19-biblio}

\end{document}